\documentclass{aa}  
\usepackage[authoryear]{natbib}
\usepackage{graphicx}
\usepackage{supertabular} 
\usepackage{txfonts}
\begin{document}
   \title{Origin of Chromatic Features in Multiple Quasars}

   \subtitle{Variability, Dust, or Microlensing}

   \author{Atsunori Yonehara\inst{1,2}\fnmsep\thanks{JSPS Postdoctoral Fellowships for Research Abroad} 
          \and
          Hiroyuki Hirashita\inst{3}
          \and
          Philipp Richter\inst{4,5}
          }

   \offprints{A. Yonehara}

   \institute{Astronomisches Rechen-Institut, 
              Zentrum f\"{u}r Astronomie der Universit\"{a}t Heidelberg, 
              M\"{o}nchhofstra{\ss}e 12-14, Heidelberg, 69120, Germany
         \and
              Department of Physics, Faculty of Science, 
              Kyoto Sangyo University, 
              Motoyama, Kamigamo, Kita-ku, Kyoto, 603-8555, Japan \\
              \email{yonehara@cc.kyoto-su.ac.jp} 
         \and 
              Center for Computational Science, University of Tsukuba, 
              Tennodai 1-1-1, Tsukuba, Ibaraki 305-8577, Japan \\
              \email{hirasita@ccs.tsukuba.ac.jp}
         \and
              Argelander-Institut f\"{u}r Astronomie, Universit\"{a}t Bonn, 
              Auf dem H\"{u}gel 71, 53121, Bonn, Germany
         \and 
              Institut f\"{u}r Physik, Universit\"{a}t Potsdam, 
              Am Neuen Palais 10, 14469, Potsdam, Germany \\
              \email{prichter@astro.physik.uni-potsdam.de}            
             }

   \date{Received ; accepted }

 
  \abstract
   {}
   {In some of the lensed quasars, color differences between
    multiple images are observed at optical/near-infrared wavelengths. 
    There are three possible origins of the color differences:
    intrinsic variabilities of quasars,  
    differential dust extinction, and quasar microlensing. 
    We examine how these three possible scenarios can reproduce 
    the observed chromaticity.}
   {We evaluate how much color difference 
    between multiple images can be reproduced 
    by the above three possible scenarios with realistic models; 
    (i) an empirical relation for intrinsic variabilities of quasars, 
    (ii) empirical relations for dust extinction and theoretically 
    predicted inhomogeneity in galaxies, or 
    (iii) a theoretical model for quasar accretion disks and 
    magnification patterns in the vicinity of caustics.}
   {We find that intrinsic variabilities of quasars 
    cannot be a dominant source responsible for  
    observed chromatic features in multiple quasars. 
    In contrast, either dust extinction or quasar microlensing 
    can nicely reproduce the observed color differences 
    between multiple images in most of the lensed quasars.
    Taking into account the time interval between 
    observations at different wavebands in our estimations, 
    quasar microlensing is a more realistic scenario 
    to reproduce the observed color differences than dust extinction.  
    All the observed color differences presented in this paper 
    can be explained by a combination of these two effects, 
    but monitoring observations at multiple wavebands are necessary
    to disentangle these.}
   {}

   \keywords{Accretion, accretion disks -- h
             Gravitational lensing --
             dust, extinction --
             quasars: general}

   \maketitle
%

\section{Introduction}

Gravitational lensing represents a major observational tool 
in modern astronomy and astrophysics. 
Although its practical application has been mainly focused on 
cosmological aspects at the beginning \citep[e.g.,][]{zel,refsdal,pacz}, 
the subjects extend to studies of galactic structure 
\citep[e.g.,][]{oguri}, searches for exo-planets including 
earth mass planets \citep[e.g.,][]{beaulieu} and others.  
A fundamental feature of gravitational lensing is achromaticity; 
this means that in principle gravitational lensing effects 
show no dependence on wavelength. 
In many situations, this achromaticity is important to 
discriminate gravitational lensing effects from 
other annoying phenomena for the detection; 
for instance, variable stars in galactic microlensing surveys 
\citep[e.g.,][]{alcock}.  
However, some gravitational lensing phenomena are associated 
with unexpected chromatic features; 
that is, observed properties have wavelength dependence 
\citep[e.g.,][]{falco}.  
In such cases, the gravitational lensing hypothesis 
has been confirmed using other supporting observations; 
studying these objects is important to obtain 
deeper insight into the effect of chromaticity in lensed objects.  

An important example appears as 
the color difference 
\footnote{Throughout this paper, we use the word `color difference' 
as difference between colors of multiple images, 
and never use it as difference between different colors.} 
between multiple images of lensed quasars.  
Although multiple images of lensed quasars should have 
the same color as the non-lensed image 
according to the principle of gravitational lensing,  
some lensed quasars show a clear color difference between different images. 
One possible explanation for the observed chromaticity is 
differential dust extinction inside the lens galaxy.  
It is apparent that the spatial distribution of interstellar
gas and dust is inhomogeneous in galaxies.   
The light path of different images pass different parts of  
the lens galaxy, and it is natural to consider that 
different images are affected by spatially varying 
dust extinction characteristics. 
Within this scenario, the amount of color difference between images 
at given wavebands is mainly determined by three quantities:  
the redshift of the lens galaxy, the column density of dust, 
and the extinction properties of dust. 
Thus, we can probe these three quantities 
from photometric data of more than three wavebands. 
\citet{falco} have investigated a method to probe 
the dust extinction in distant galaxies, 
\citet{toft, munoz, elias, mediv} have 
explored the extinction law at high-redshift galaxies, 
and \citet{dai} have probed the dust-to-gas ratio 
of galaxies at cosmological distance 
by combining the hydrogen column density obtained with 
the Chandra X-ray Observatory and the color excess obtained 
by \citet{falco}. 
This enables us to measure the absorption properties 
at distant galaxies directly, and it is complement to 
emission-weighted measurements of dust properties.  
In contrast, \citet{jean} have proposed a method to 
estimate the redshift of the lens galaxy under
an assumption on the extinction properties. 
Direct detection of the emission from the lens galaxy
is not necessary for this method, 
and it can be a strong tool to estimate redshift of 
faint or so-called ``dark'' lens galaxies.  
  
Unfortunately, differential extinction is not the only scenario 
able to explain the observed chromaticity in multiple quasars. 
Because multiple images have different light paths, 
a delay in the arrival time is always present between the images. 
This indicates that we observe slightly different epochs  
of the lensed quasar at the same time via multiple images. 
Quasars intrinsically change not only their luminosity 
but also their color with time \citep[e.g.,][]{wilhite}. 
Incorporating time delay and intrinsic color variabilities, 
one could explain the observed chromaticity to some degree. 
In addition, quasar microlensing can be yet another candidate to 
explain the observed chromaticity. 
A theoretical investigation of chromatic features of quasar microlensing 
has initially been made by \citet{wam91} 
in the case of so-called ``Huchra's lens (Q2237+0305)''.  
In other systems, possible evidence for chromaticity due to 
microlensing has been reported, for example, by \citet{burud02a, nakos}.     
Since there is no essential difference 
between Huchra's lens and other multiple quasars,  
color changes due to quasar microlensing can be 
an origin of the observed chromaticity and can work as 
a contaminant in exploring dust extinction in the lens galaxy. 

In summary, there are three possibilities for 
the chromaticity in a multiple quasar: 
(i) intrinsic quasar variability, 
(ii) differential dust extinction, 
and (iii) quasar microlensing. 
In this paper, we consider all these three possible scenarios  
to explain the observed chromaticity  
with realistic theoretical models and reliable empirical relations. 
We will briefly introduce the observed chromaticity in section 2. 
The three possible scenarios for the observed chromaticity 
are individually examined in sections 3, 4, and 5.   
Discussions about possible and realistic origin of  
the observed chromaticity are presented in section 6.     
Throughout this paper, we adopt the following cosmological parameters: 
$\Omega_{\rm M}=0.3$, $\Omega_{\rm \Lambda}=0.7$, 
and $H_0=70~{\rm km s^{-1} Mpc^{-1}}$ 
\citep[][]{spergel}.

\section{Observed Chromaticity}

Several observational programs for lensed quasars 
are currently carried out, 
and we here use the data taken by one of such programs, CASTLES 
\footnote{CfA-Arizona Space Telescope LEns Survey. 
The data and details of this program are presented in 
{\tt http://www.cfa.harvard.edu/castles/} . 
See also publications of CASLTES \citep[e.g.,][]{lehar} for the details.}. 
CASTLES provides us with photometric data of lensed quasars of 
high and equal quality, and the data are suitable for 
comparison with theoretical models and predictions. 
Roughly 100 lensed quasars have been listed so far,  
but we pick up only 25 objects whose photometric data 
at F160W of {\it Hubble Space Telescope} ({\it HST}) NICMOS, 
and at F555W and F814W of {\it HST} WFPC2 are 
publically available online \footnote{until December, 2006}. 
This sample enables us to use two independent colors 
to disentangle the origin of the observed chromaticity. 
This is useful for breaking some degeneracy 
which will easily occur from a single color information, 
for example, degeneracy between extinction properties of dust 
and amount of dust.
Details of the sample lensed quasars and lens galaxies are 
summarized in Tables~\ref{tab:obj}, and \ref{tab:gal}. 

Here, we briefly summarize how the photometric data are
analyzed and obtained by \cite{lehar}.  
The brightness of objects was measured by $\chi^2$ minimization 
among the observed images and models for lensed quasars and lens galaxies.
A lensed quasar image is modeled by a PSF with
a sharp central core (FWHM $\sim 0.1~{\rm arcsec}$) and
an outer skirt of low intensity ($\le 1\%$ of the peak intensity),
and lens galaxies are modeled by an ellipsoidal exponential disk
or de Vaucouleurs models. 
Each lensed quasar image has 3 parameters for fitting 
(2-dimensional position and flux), and each lens has
6 parameters for fitting (2-dimensional position, flux,
major axis, axis ratio, position angle of major axis). 
Uncertainties of these decomposition procedures are 
included in errors presented in Tables~\ref{tab:obj}, and \ref{tab:gal}.


\addtocounter{table}{1}

\begin{table*}
\caption{Summary of lens galaxies in the current sample. 
Their magnitude and error at F160W, at F555W and at F814W are presented. 
All the data are taken from CASTLES web page. 
Types of the lens galaxies indicated by their spectrum, color or 
brightness profile are also shown. 
Early and late type is presented by `E' and `L', respectively. 
For a sample with two lens galaxies, they are denoted by G and G'.
References. 
(1) \citealt{lehar}; (2) \citealt{tonry}; (3) \citealt{fass98}; 
(4) \citealt{keeton}; (5) \citealt{burud98}; (6) \citealt{lubin}; 
(7) \citealt{young}; (8) \citealt{ango}; (9) \citealt{kochan}; 
(10) \citealt{remy}; (11) \citealt{impey98}; (12) \citealt{impey96}; 
(13) \citealt{chav}; (14) \citealt{koop02}; (15) \citealt{fass99};  
(16) \citealt{lopez}; (17) \citealt{huchra}. }  
\label{tab:gal}
\centering
\begin{tabular}{r c c c c c c c}
\hline \hline
 Object Name & $m_{F160W}$ & $m_{F555W}$ & $m_{F814W}$ & 
 \multicolumn{3}{c}{Galaxy Type} \\
 ~ & ~ & ~ & ~ & Spectrum & Color & Profile \\
\hline 
Q0142-100~G     & $16.63 \pm 0.03$ & $20.81 \pm 0.02$ & $18.72 \pm 0.05$ & - & E(1) & E(1) \\
B0218+357~G     & $17.50 \pm 0.04$ & $21.95 \pm 0.24$ & $20.06 \pm 0.14$ & - & L(1), E(1) & L(1) \\
MG0414+0534~G   & $17.54 \pm 0.14$ & $24.17 \pm 0.15$ & $20.91 \pm 0.05$ & - & E(2) & - \\
B0712+472~G     & $17.16 \pm 0.15$ & $21.75 \pm 0.10$ & $19.56 \pm 0.07$ & E(3) & E(4) & - \\
RXJ0911+0551~G  & $17.93 \pm 0.08$ & $22.97 \pm 0.22$ & $20.47 \pm 0.09$ & - & - & E(5), L(5) \\
SBS0909+523~G   & $16.75 \pm 0.74$ & $18.29 \pm 0.55$ & $17.12 \pm 1.12$ & E(6) & E(1) & - \\ 
BRI0952-0115~G  & $18.95 \pm 0.16$ & $23.67 \pm 0.08$ & $21.21 \pm 0.04$ & - & E(4) & E(1) \\
Q0957+561~G     & $15.14 \pm 0.09$ & $19.05 \pm 0.06$ & $17.12 \pm 0.03$ & - & E(4) & E(7) \\
         ~G'    & $17.92 \pm 0.04$ & $21.87 \pm 0.25$ & $19.99 \pm 0.24$ & E(8) & - & - \\
LBQS1009-0252~G & $19.30 \pm 0.12$ & $24.05 \pm 0.54$ & $21.99 \pm 0.04$ & - & - & E(1) \\
B1030+071~G     & $17.64 \pm 0.15$ & $22.71 \pm 0.12$ & $20.24 \pm 0.13$ & E(3) & E(1) & E(1) \\
         ~G'    & $19.11 \pm 0.08$ & $24.56 \pm 0.20$ & $22.04 \pm 0.13$ & - & E(1) & L(1)\\
HE1104-1805~G   & $17.47 \pm 0.27$ & $23.14 \pm 0.58$ & $20.01 \pm 0.10$ & - & E(9,10) & - \\
PG1115+080~G    & $16.66 \pm 0.04$ & $20.74 \pm 0.04$ & $18.92 \pm 0.02$ & - & E(4) & E(11) \\
B1422+231~G     & $17.57 \pm 0.20$ & $21.80 \pm 0.17$ & $19.66 \pm 0.25$ & - & E(12) & - \\
SBS1520+530~G   & $17.84 \pm 0.06$ & $21.96 \pm 1.24$ & $20.16 \pm 0.13$ & - & L(13) & - \\
B1600+434~G     & $18.30 \pm 0.13$ & ~---~            & $20.78 \pm 0.41$ & L(4) & - & - \\
MG2016+112~G    & $18.46 \pm 0.09$ & $25.12 \pm 1.06$ & $21.95 \pm 0.09$ & - & - & E(14) \\
          ~G'   & $23.08 \pm 0.62$ & $24.92 \pm 0.30$ & $24.56 \pm 0.09$ & - & - & - \\
B2045+265~G     & $18.25 \pm 0.26$ & $23.86 \pm 0.22$ & $21.15 \pm 0.19$ & L(15) & - & - \\
HE2149-2745~G   & $17.61 \pm 0.10$ & $21.18 \pm 0.09$ & $19.56 \pm 0.03$ & - & - & E(16)\\
Q2237+0305~G    & $12.22 \pm 0.22$ & $15.49 \pm 0.22$ & $14.15 \pm 0.20$ & E(17) & - & L(17) \\
\hline
QJ0158-4325~G   & $19.75 \pm 0.27$ & $22.96 \pm 0.22$ & $22.42 \pm 0.83$ & - & - & - \\
APM08279+5255~G & $15.11 \pm 0.04$ & $18.70 \pm 0.07$ & $16.93 \pm 0.05$ & - & - & - \\
FBQ0951+2635~G  & $17.86 \pm 0.23$ & $21.02 \pm 0.20$ & $19.67 \pm 0.23$ & - & - & - \\
            ~G' & $18.50 \pm 0.30$ & $20.06 \pm 0.38$ & $20.33 \pm 0.17$ & - & - & - \\
Q1017-207~G     & $19.26 \pm 0.06$ & $25.48 \pm 0.73$ & $21.82 \pm 0.48$ & - & E(9) & E(1) \\
Q1208+101~G     & ~---~            & ~---~            & ~---~            & - & - & - \\
FBQ1633+3134~G  & $16.80 \pm 1.10$ & $19.21 \pm 3.38$ & $18.26 \pm 0.89$ & - & - & - \\
\hline
\end{tabular}
\end{table*}

\begin{figure}
\centering
\includegraphics[width=\hsize]{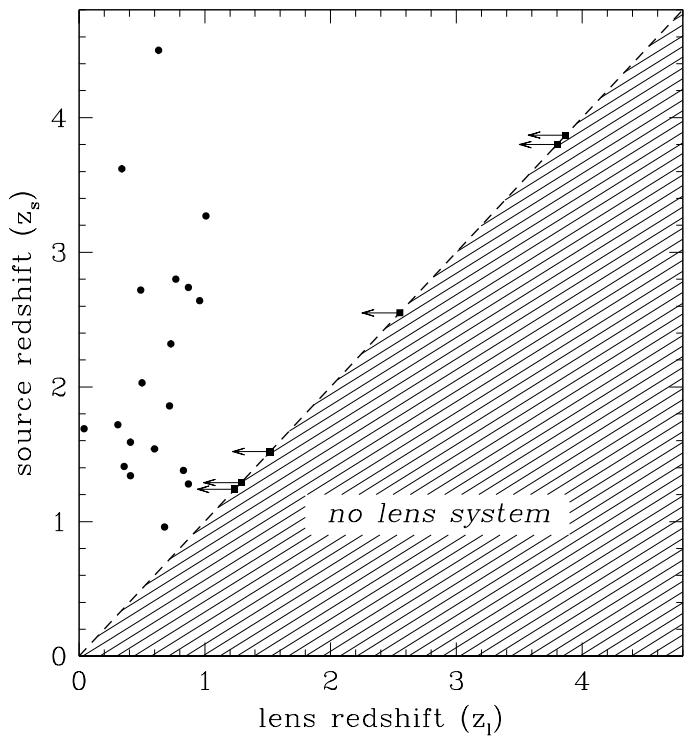}
\caption{The lens redshift ($z_l$) and the source redshift ($z_s$) 
of the current samples are presented by the filled circles in this figure. 
Objects without the lens redshift are indicated by 
the filled squares with arrows.}
\label{fig:redshift}
\end{figure}

The redshifts of the lens ($z_l$) and the source ($z_s$) 
listed in Tables~\ref{tab:obj} are also taken from CASTLES, 
and presented in Figure~\ref{fig:redshift}. 
The redshifts of all the lensed quasars are known as presented 
in Tables~\ref{tab:obj}. 
In contrast, the redshifts of the lens galaxies are not known 
in all cases.  
Objects with unknown redshifts or with tentative redshifts   
are listed at the bottom of Table~\ref{tab:obj}. 
For these objects, the upper limits of their lens redshifts   
are set to be the source redshifts 
and are denoted by the filled squares with arrows  
in Figure~\ref{fig:redshift}. 
The source quasars and the lens galaxies are distributed 
at $z \simeq 1$--$4$ and at $z \simeq 0$--$1$, respectively
\footnote{\cite{eigen} has recently measured that 
the lens redshifts of FBQ0951+2635 and HE2149-2745 
are $0.260$ and $0.603$, respectively. 
However, these redshifts are within this redshift range 
of the lens galaxies and our conclusion are not modified 
by the measurements.}. 
In total, now we have 25 objects (or 40 image pairs) 
for our current purpose. 

\begin{figure*}
\centering
\includegraphics[width=\textwidth]{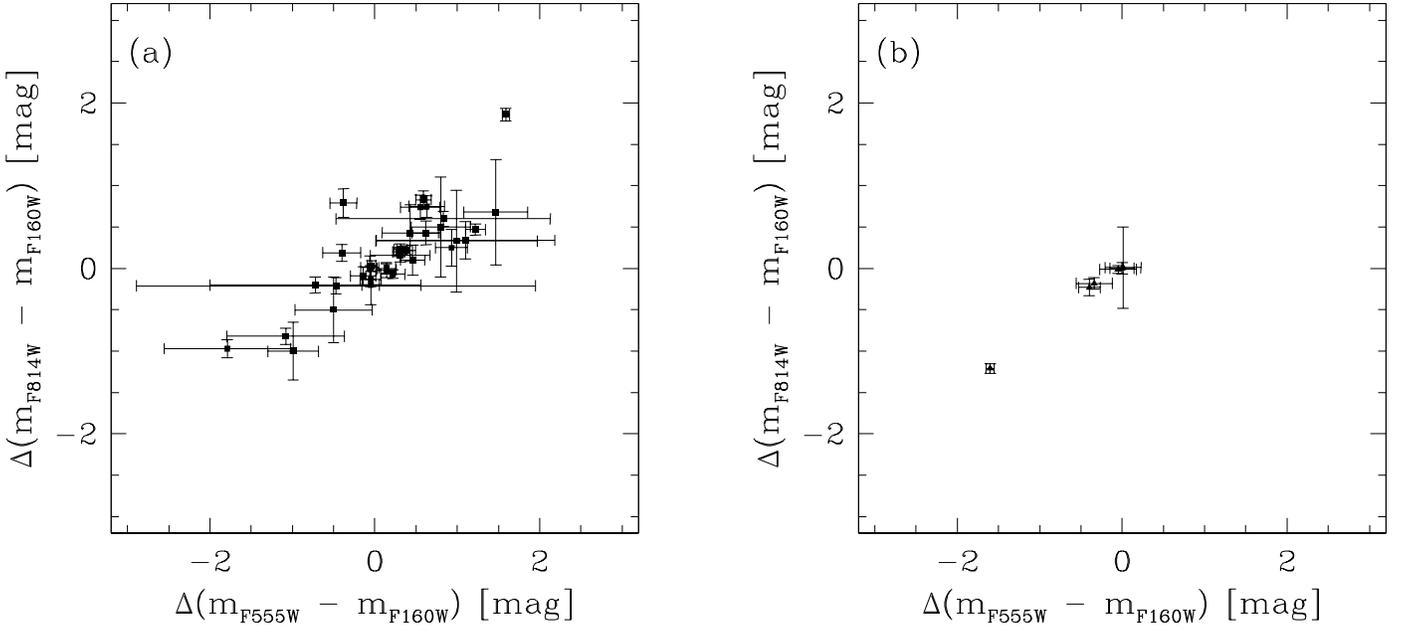}
\caption{Color differences between multiple images 
are plotted with error bars. 
For each system, the brightest image in F160W 
has been used as color reference.  
The abscissa and ordinate represent the color difference derived from 
F555W and F160W ($\Delta (m_{F555W} - m_{F160W})$), and 
that from F814W and F160W ($\Delta (m_{F814W} - m_{F160W})$), respectively. 
Panel (a) is for lensed quasars 
for which the redshift of the lens galaxy is known, 
while Panel (b) is for lensed quasars 
for which the redshift of the lens galaxy is unknown.}
\label{fig:color}
\end{figure*}

In Table~\ref{tab:obj}, the colors of each lensed quasar image 
for F555W -- F160W and F814W -- F160W, and the F160W magnitudes  
of each image are also presented together with their errors.   
The errors are evaluated in the usual manner, 
namely by calculating the square root of the sum of 
the squares of the individual error.  
It is apparent from Table~\ref{tab:obj} that some of the lensed quasars 
show color differences between different images.  
This feature can also be found more clearly in Figure~\ref{fig:color}; 
the color differences of multiple images   
relative to that of the brightest image at F160W, 
$\Delta (m_{F555W} - m_{F160W})$ and $\Delta (m_{F814W} - m_{F160W})$. 
Images located at the lower left part in Figure~\ref{fig:color}  
are relatively bluer than the brightest image at F160W in the system. 
Gravitational lensing effect has no wavelength dependence in principle, 
and the magnification factor should be the same at different wavebands.  
Since the magnitude differences between images correspond to 
the ratio of magnification factors only if the macro lensing
contributes to the magnification, 
the magnitude difference should be identical among any wavebands.
That is, observational data points should be located at the origin, 
$(0,0)$, in Figure~\ref{fig:color} within the error. 
Though some of the multiple quasars show achromaticity as expected, 
others show anomalous chromaticity; 
i.e., some data points in Figure~\ref{fig:color} significantly 
deviate from the origin, ranging up to $\sim 2~{\rm mag}$. 

Although there is a wavelength overlap among 
these two combinations of the color differences,  
F555W -- F160W and F814W -- F160W, 
we choose these combinations because of their small errors 
compared to combinations of F555W -- F814W. 
The lower-left to upper-right trend of observational data points 
in Figure~\ref{fig:color} or following similar figures 
is a natural consequence of the wavelength overlap, 
and hereafter we just focus on deviations of observational data 
from the expected values of our calculations rather 
than the trend. 

Here, we review what kind of effect should be included 
for interpretation of the observed chromaticity. 
By using the intrinsic magnitude of a quasar ($m_{\rm int}$), 
the expected magnitude of an image of multiple quasars ($m_{\rm obs}$)  
at wavelength ($\lambda$) at any given time ($t$) is 
written as follows:  
\begin{equation}
m_{\rm obs}(\lambda,t) = m_{\rm int}(\lambda,t-\tau) 
 + A\left( \frac{\lambda}{1+z_l} \right) 
  - 2.5 \log \left[ \mu(\lambda, t) \right],  
\label{eq:basic}
\end{equation}  
where $\tau$, $A$, and $\mu$ indicate 
the arrival time delay due to gravitational lensing, 
the extinction along a path from the source to observer, 
and the magnification due to gravitational lensing, respectively.  
To include the finite source size effect, 
as is different from a usual expression, 
the magnification due to gravitational lensing is presented 
as a function of wavelength and time in equation~\ref{eq:basic}   
(see section 5 for more details). 
These three terms all affect the multiple images with
different magnitudes for different images.   
Thus, they can be possible origins of the observed chromaticity. 
In the following three sections, 
we examine how these effects may produce the observed chromaticity.

\section{Intrinsic Variabilities of Quasars}

It is well known that quasars show 
temporal flux variabilities at any wavebands, 
and the color becomes bluer when the flux gets brighter 
\citep[e.g.,][]{wilhite, cri97}.
This means that the color of quasars changes with time
according to the flux variabilities.   
Additionally, the arrival time delay 
due to gravitational lensing is different at different images.  
This produces the relative time delay between multiple images 
of the same quasar, and multiple images observed at the same epoch  
correspond to images at intrinsically different epochs.
Therefore, it is probable that 
multiple images observed at the same time show different colors. 
To estimate this effect, we evaluate typical time delays between images 
and typical color change of quasars for a given time interval. 

Though the actual time delay is determined by 
the density profile of the lens galaxy, 
the location of the source respective to the lens on the sky, 
the source redshift, and the lens redshift, 
we can roughly evaluate the typical time delay 
from observed properties of multiple quasars. 
By assuming that the density profile of a lens galaxy is 
approximated by a singular isothermal sphere \citep[SIS; e.g.,][]{sef}, 
the time delay between two images, 
one image with positive parity 
(located at $\theta_+$ from the lens galaxy) 
and another image with negative parity 
(located at $\theta_-$ from the lens galaxy),  
is expressed by 
\begin{eqnarray}
 \tau &=& \frac{1+z_l}{2c} \frac{D_{ol}D_{os}}{D_{ls}} 
  \left( \theta_+^2 - \theta_-^2 \right)  \nonumber \\
      &=& 30 \left( \frac{1+z_l}{2} \right) 
       \left( \frac{D_{ol}D_{os}/D_{ls}}{1~{\rm Gpc}} \right) 
        \left( \frac{\theta_+^2 - \theta_-^2}{1~{\rm arcsec}^2} \right) 
         ~{\rm days}, 
\end{eqnarray}
where $D_{ol}$, $D_{os}$, and $D_{ls}$ represent 
the angular diameter distances from observer to the lens, 
from observer to the source, and from the lens to the source, respectively. 
Since the image separation is on the order of $1~{\rm arcsec}$ 
in most multiple quasars, we can roughly estimate
$\theta_+^2 - \theta_-^2$ to be $\sim 1~{\rm arcsec}^2$.  
Thus, we adopt $30~{\rm days}$ for a typical time delay 
between multiple images. 

Further, by using recent studies of intrinsic variabilities of quasars, 
we can also evaluate the color change of quasars on any given timescale. 
\citet{vanden, ivezic} have investigated how quasar variabilities in 
the rest-frame optical/UV regime depend on other observational quantities 
such as rest-frame time lag. 
They analyzed relations among these quantities and the structure function 
\citetext{a commonly used statistical measure for variabilities; 
e.g., \citealp{gucchi}}
by using imaging data of quasars obtained by Sloan Digital Sky Survey.   
The resulting structure function ($V$) is expressed as  
\begin{equation}
 V = \left( 1 + 0.024 M_i \right) 
   \left(  \frac{\Delta t_{\rm RF}}{\lambda_{\rm RF}} \right)^{0.3}
     ~{\rm mag}, 
\label{eq:qsosf}
\end{equation}
where $M_i$, $\Delta t_{\rm RF}$, and $\lambda_{\rm RF}$ represent 
the $i$ band absolute magnitude of a quasar in units of magnitude, 
the rest-frame time lag between observations in units of days, 
and the rest-frame wavelength in units of \AA, respectively. 
Equation~\ref{eq:qsosf} indicates that a larger flux change of quasars 
occurs at fainter quasars, at longer time lag, and/or at shorter wavelength. 
The absolute $i$ band magnitude of the quasar sample 
is distributed from $\sim -21$ to $\sim -30$ \citep[][]{vanden}.  
Since $\Delta t_{\rm RF}$ and $\lambda_{\rm RF}$ have the same 
redshift dependence, both of these quantities can be 
replaced by the observer-frame quantities. 
 
Substituting the typical time delay above (30 days)   
and the effective wavelength of each filter into equation~\ref{eq:qsosf}, 
we obtain the expected flux variation of quasars.  
The expected values for magnitude difference ($\delta m$)   
with a time lag of $30~{\rm days}$ is 0.043 -- 0.075~mag, 
0.059 -- 0.105~mag, and 0.053 -- 0.093~mag  
at F160W (mean wavelength: $16071$~\AA), 
F555W (mean wavelength: $5337$~\AA), and 
F814W (mean wavelength: $7900$~\AA), respectively. 
Here, the maximum and the minimum value of magnitude differences 
are obtained for $M_i=-21$ and $M_i=-30$, respectively.  
Even if we take into account filter responses, 
these values would not change much.   
The wavelength coverage of all the filters is 
less than $\sim \pm20\%$ of the wavelength center 
(see also the upper panel of Figure~\ref{fig:extinction}), 
and the expected flux variations are less than 
$\pm7\%$ from the above values within the coverage 
(see equation~\ref{eq:qsosf}). 
By comparing these expected values for magnitude difference 
at different wavebands, we are able to estimate 
the possible range of the expected color differences.  

In general, the range of the expected color difference 
between two images for two bands labeled with $j$ and $k$, 
$|\Delta (m_j - m_k)|$, is estimated by 
$| |\delta m_j| - |\delta m_k| | < |\Delta (m_j - m_k)| 
 < |\delta m_j| + |\delta m_k|$. 
The maximum value corresponds to the case where   
variabilities in band $j$ and that in band $k$ 
have completely negative correlation, 
i.e., when a quasar gets brighter in band $j$, 
 a quasar always gets fainter in a $k$ band. 
The minimum value corresponds to the case where  
variabilities in these two bands have completely positive correlation, 
i.e., when a quasar becomes brighter in band $j$, 
 it always gets brighter also in band $k$.    
For quasars with $M_i=-21$, the fainter end of 
the sample of \citet{vanden}, 
the expected color differences are ranging  
from $|0.105-0.075|=0.030$~mag to $0.105+0.075=0.180$~mag  
for $| \Delta (m_{\rm F555W} - m_{\rm F160W}) |$,  
and from $|0.093-0.075|=0.018$~mag to $ 0.093+0.075=0.168$~mag  
for $| \Delta (m_{\rm F814W} - m_{\rm F160W}) |$. 
Since the observed chromaticities as shown in Figure~\ref{fig:color} 
have values up to $\sim 2~{\rm mag}$, 
it is impossible to reproduce all the observed chromaticity 
only by this scenario which is able to reproduce 
a color difference up to $0.2~{\rm mag}$. 
For intrinsically brighter quasars, 
the expected flux change within a time interval is rather small, 
and the expected color differences are smaller than the observed values. 
Moreover, quasar variabilities at any two wavebands presumably 
have positive correlation with each other rather than a negative correlation 
\citep[e.g.,][]{wilhite}, and the actual color difference 
should be smaller than our estimated maximum values.  

Of course, such positive correlation will be lost 
if observations for an object at different wavebands  
are carried out at different times. 
Denoting this observational interval among 
two bands (labeled with $j$ and $k$) as $t_{lag}$, 
the color difference including time delay ($\tau$) 
is expressed as 
$\Delta (m_j - m_k) = \left[ m_j (t) - m_k(t + t_{lag}) \right] 
 - \left[ m_j(t + \tau) - m_k(t + \tau + t_{lag}) \right]  
= \left[ m_j (t) - m_j(t + \tau) \right] 
 - \left[ m_k(t + t_{lag}) - m_k(t + \tau + t_{lag}) \right]$. 
This relation indicates that the expected color differences 
are a simple combination of magnitude differences at each band  
adopted above. 
Since an actual time delay is nothing to do with 
the observational interval among different wavebands, 
$t_{lag}$ is the only factor that makes difference from 
the previous situation, i.e., without the observational interval.    
As presented by \citet{lehar}, 
the observational interval among different wavebands 
spanned up to $t_{lag} \sim 2$ years for some objects,  
and it is comparable to maximum timescale of quasar variabilities  
\citep[e.g.,][]{ivezic} 
\footnote{Generally, structure function of intrinsic quasar variability  
consist of two parts; power-law component below a certain timescale 
and flat component above the timescale. 
Since the flat component of the structure function  
originates from only a random process, 
the timescale where the power-law component ends 
corresponds to maximum timescale of quasar variability 
due to some physical process.}.  
Since physical correlation could be lost 
after a timescale longer than maximum timescale of quasar variability,   
a completely negative correlation among two wavebands 
is even possible when the observational interval is comparable to 
or longer than $\sim 2$ years. 
However, even if it is the case,  
the amplitude of variations estimated above fixes 
upper bounds on the expected color differences, and 
it is impossible to produce color differences larger than 
$| \Delta (m_{\rm F555W} - m_{\rm F160W}) | = 0.180 {\rm ~mag}$ and 
$| \Delta (m_{\rm F814W} - m_{\rm F160W}) | = 0.168 {\rm ~mag}$.  
Thus, the expected color differences 
should be still within the range estimated above, 
and thus the observed color differences cannot be explained 
by intrinsic variabilities of quasars alone.  


The model for the lens galaxies that we used here for estimating 
the time delay between multiple images is rather simple, 
though the time delay depends on the applied lens model. 
This has already been mentioned in previous studies \citep[e.g.,][]{oguri}, 
and the expected time delay of more realistic lens models can be 
reduced down to an order of magnitude from that of SIS. 
Diversity of the expected time delay is also investigated by 
different approaches \citep[e.g.,][]{saha06}, and the time delay 
range would be typically between $\sim 10$ and $\sim 300$ days. 
Further, if the lens redshift is unknown, 
the expected time delay of any lens model can have 
a dispersion of one order of magnitude for a given image separation. 
Consequently, taking a more realistic lens model into account,  
the expected time delay can change up to two orders of magnitude. 
Even if this is the case, the expected color difference 
between multiple images can change only by a factor of 4 
(see equation~\ref{eq:qsosf}). 

From the observational point of view, 
the measured time delay shows wide variety as predicted 
by the previous theoretical studies \citep[e.g.,][]{oguri,saha06}, 
and the measured values are different from $30~{\rm days}$. 
In Table~\ref{tab:tdelay}, we summarize the time delay 
between multiple images in the current sample. 
The time delay of objects which are not listed in Table~\ref{tab:tdelay}, 
Q0142-100, MG0414+0534, B0712+472, SBS0909+523, LBQS1009-0252, B1030+071, 
MG2016+112, B2045+265, Q2237+0305, QJ0158-4325, APM08279+5255, BRI0952-0115, 
Q1017-207, Q1208+101, and FBQ1633+3134, 
has not been successfully measured yet.
As we can see in Table~\ref{tab:tdelay},  
some multiple quasars have one order of magnitude longer time delay 
than the typical value applied here \citep[e.g.,][]{kundic}.  
In such systems, of course, the expected flux change is 
larger than that we estimated for multiple quasars with 
the time delay of $30~{\rm days}$.  
However, the color difference becomes only twice the value 
estimated above (see equation~\ref{eq:qsosf}), 
and the observed color differences are not still explained. 
Again, it is clear that intrinsic variabilities of quasars alone 
cannot reproduce observed color difference between multiple images.

\begin{table*}
\caption{Time delay between multiple images in current sample 
is presented in units of day. 
Wavebands used for the measurement are also presented with the reference. 
Only objects with the available time delay are listed here.
References. (1) \citealt{biggs}; (2) \citealt{hjorth}; 
(3) \citealt{ullan}; (4) \citealt{kundic};  (5) \citealt{haarsma}; 
(6) \citealt{ofek}; (7) \citealt{schec97}; (8) \citealt{patnaik}; 
(9) \citealt{burud02b}; (10) \citealt{koop00}; (11) \citealt{burud02a}; 
(12) \citealt{jakob}.}    
\label{tab:tdelay}
\centering
\begin{tabular}{r | c c | c c}
\hline \hline
Object and Image Pair & Optical Delay & Waveband &
 Radio Delay & Waveband \\
  ~ & (day) & ~ & (day) & ~ \\
\hline
B0218+357~A--B     & ~---            & ~                  & $10.5 \pm 0.4$    & 8.4 and 15~GHz (1) \\
RXJ0911+0551~A--B  & $146 \pm 8$     & I-band (2)         & ~---              & ~ \\
SBS0909+523~A--B   & $45^{+1}_{-11}$ & R-band (3)         & ~---              & ~ \\
Q0957+561~A--B     & $417^{+3}_{-3}$ & g-band (4)         & $459^{+12}_{-15}$ & 6~cm (5) \\
                 ~ & $420^{+6}_{-9}$ & r-band (4)         & $397^{+12}_{-12}$ & 4~cm (5) \\
HE1104-1805~A--B   & $161^{+7}_{-7}$ & V- and R-band (6)  & ~---              & ~ \\
PG1115+080~A--C    & $9.4$           & V-band (7)         & ~---              & ~ \\
          ~C--B    & $23.7 \pm 3.4$  & V-band (7)         & ~---              & ~ \\
B1422+231~A--B     & ~---            & ~                  & $1.5 \pm 1.4$     & 8.4 and 15~GHz (8) \\
         ~A--C     & ~---            & ~                  & $7.6 \pm 2.5$     & 8.4 and 15~GHz (8) \\
         ~B--C     & ~---            & ~                  & $8.2 \pm 2.0$     & 8.4 and 15~GHz (8) \\
SBS1520+530~A--B   & $130 \pm 3$     & R-band (9)         & ~---              & ~ \\
B1600+434~A--B     & ~---            & ~                  & $47^{+5}_{-6}$    & 8.5~GHz (10) \\
HE2149-2745~A--B   & $103 \pm 12$    & V- and i-band (11) & ~---              & ~ \\
\hline
FBQ0951+2635~A--B  & $16 \pm 2$      & R-band (12)        & ~---              & ~ \\
\hline 
\end{tabular}
\end{table*}

\section{Differential Dust Extinction}

The light paths of different images intersect 
different positions of the lens galaxy. 
In general, the column density of gas and dust along different lines of sight 
differs in all galaxies, and different images experience 
different levels of dust extinction. 
We can also consider that the shape of extinction curve is 
different at different positions in the lens galaxy. 
For simplicity, however, we basically take into account only 
the inhomogeneity of dust column densities, 
but we also examine the difference among 
various empirical extinction curves derived from the Milky Way (MW) 
and the Small Magellanic Cloud (SMC).    

\begin{figure}
\centering
\includegraphics[width=\hsize]{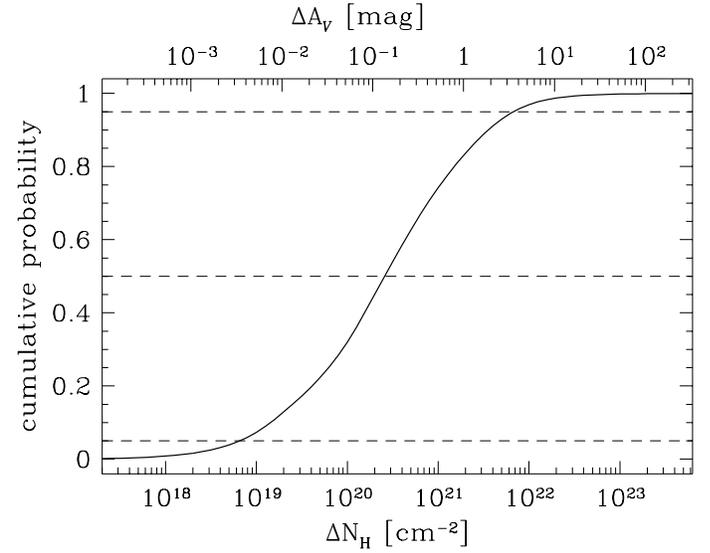}
\caption{Cumulative probability distribution of differential column densities 
of gas obtained from \citet{hirashita}. 
The lower abscissa indicates the differential column density of gas 
in units of ${\rm cm^{-2}}$. 
The corresponding differential extinction at $V$ band (in mag) 
is also denoted at the upper abscissa.  
Cumulative probabilities of $5\%$, $50\%$ and $95\%$ are 
indicated by the dashed lines.}
\label{fig:column}
\end{figure}

We here apply an extinction law derived for the MW dust 
by \citet{cardelli}.  
In this extinction law, dust extinction is characterized basically 
by two parameters; 
$A_V$, dust extinction at $V$ band which reflects amount of dust, 
and $R_V$ 
[$\equiv A_V/E_{B-V}$ ($E_{B-V}$ is the $B-V$ color excess)].  
Since $A_V$ is physically related to the column density of dust, 
we can estimate the differential dust extinction 
from differential column densities of gas by assuming values of 
the dust-to-gas ratio and $R_V$. 
In order to obtain differential column densities of gas, 
we utilize a 2-dimensional galactic-scale hydrodynamical simulation 
presented by \citet{hirashita}, 
who adopt the calculation code of \citet{wada01}.  
The result clearly shows a clumpy distribution of gas in the simulated galaxy. 
The cumulative probability distribution of differential column densities,  
$\Delta N_{\rm H}$, is shown in Figure~\ref{fig:column}; 
we randomly select $10^6$ pairs of columns 
on the 2-dimensional grid points in the simulated galaxy, 
and derive the probability distribution function of the difference of
gas column densities between each pair 
\footnote{A lensed image is usually located at the opposite direction 
from the other image with respect to the center of the lens galaxy. 
Since the gravitational evolution of clumps in galaxies 
is coherent only within a Jeans scale, 
such coherency may have a negligible effect on our result,  
i.e., spatial inhomogeneity over the galactic scale.}. 
As we can see from Figure~\ref{fig:column}, 
the expected differential column density  
from the simulation spans a large range: 
The median value is $\sim 3 \times 10^{20}~{\rm cm^{-2}}$, 
and $90\%$ of data is distributed 
from $\sim 6 \times 10^{18}~{\rm cm^{-2}}$ 
to $\sim 7 \times 10^{21}~{\rm cm^{-2}}$. 
We apply the MW dust-to-gas ratio provided by \citet{bohlin}, 
and obtain $A_V = R_V \times E_{B-V} = 3.1 \times N_{\rm H} / 
(5.8 \times 10^{21}) = 5.3 \times 10^{-22} N_{\rm H}$,
where $N_{\rm H}$ is the hydrogen column density in ${\rm cm^{-2}}$. 
In this paper, we apply the above relation 
also when we adopt values for $R_V$ other than 3.1. 
We then can convert the expected differential column densities 
into the expected differential extinction in the $V$ band, $\Delta A_V$. 
This conversion is fairly simple and 
the scale for this value is also denoted in Figure~\ref{fig:column}. 
The corresponding values of $\Delta A_V$ 
for the median, $5\%$, and $95\%$ of Figure~\ref{fig:column} are 
$\sim 0.1~{\rm mag}$, $\sim 3 \times 10^{-3}~{\rm mag}$, 
and $\sim 3~{\rm mag}$, respectively. 

\begin{figure}
\centering
\includegraphics[width=\hsize]{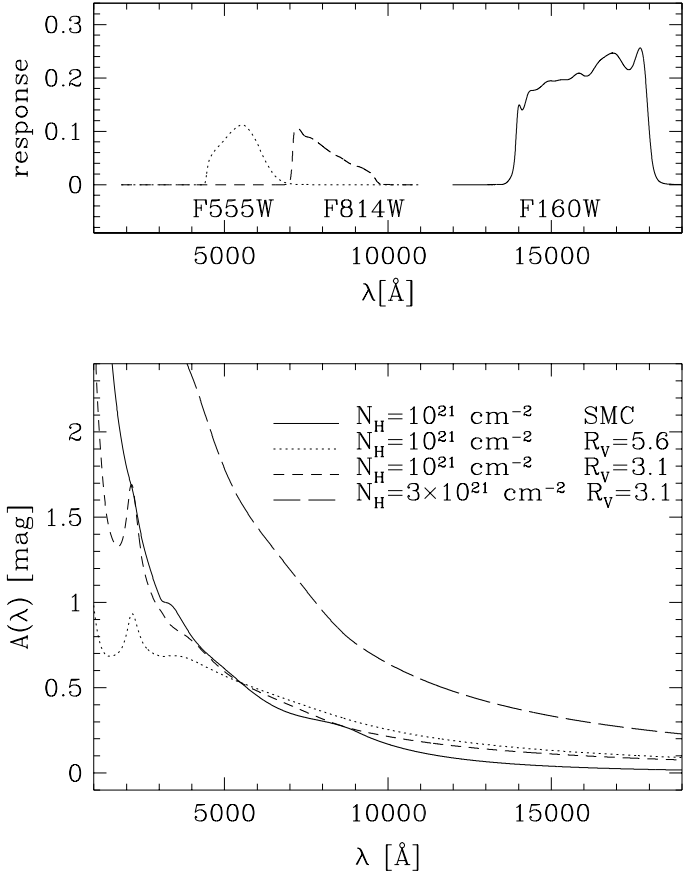}
\caption{Extinction curves for several parameter sets (lower panel) and 
filter responses (upper panel) are presented. 
In the lower panel, the solid, the dotted, the dashed, 
and the long-dashed lines 
correspond to the extinction curve of the SMC with 
$N_{\rm H}=10^{21}~{\rm cm^{-2}}$, 
that of the MW with $N_{\rm H}=10^{21}~{\rm cm^{-2}}$ and $R_V=5.6$, 
that of the MW with $N_{\rm H}=10^{21}~{\rm cm^{-2}}$ and $R_V=3.1$, and 
that of the MW with $N_{\rm H}=3 \times 10^{21}~{\rm cm^{-2}}$ 
and $R_V=3.1$, respectively. 
In the upper panel, filter responses for F555W, F814W, and F160W 
are presented by the dotted, the dashed, and the solid lines, respectively. 
The abscissa of the lower panel is the wavelength 
at the rest frame of the absorber   
(or the lens galaxy) in units of \AA, and 
that of upper panel is the wavelength at observer's frame.}
\label{fig:extinction}
\end{figure}

Additionally, it is well known that the SMC  
exhibits a very different extinction curve compared to the MW curve.
The SMC curve lacks the 2175~\AA ~bump and rises steeply toward
shorter wavelengths in the UV. 
We also adopt an SMC extinction curve in the current study. 
Several types of the SMC extinction curves have been provided,  
but we investigate the extinction law by using an observational data  
of \citet{gordon} (see appendix for details) 
which covers a relatively wide wavelength range. 
Some examples of the extinction curves 
that we adopted in this study are shown in Figure~\ref{fig:extinction}. 
Since \citet{cardelli} have presented the extinction law 
only for $R_V=2.6 - 5.6$, and since there is no guarantee that 
the empirical extinction law is valid for $R_V < 2.6$ or $R_V > 5.6$, 
we put in this paper the same limit as \citet{cardelli}.  

\begin{figure*}
\centering
\includegraphics[width=\textwidth]{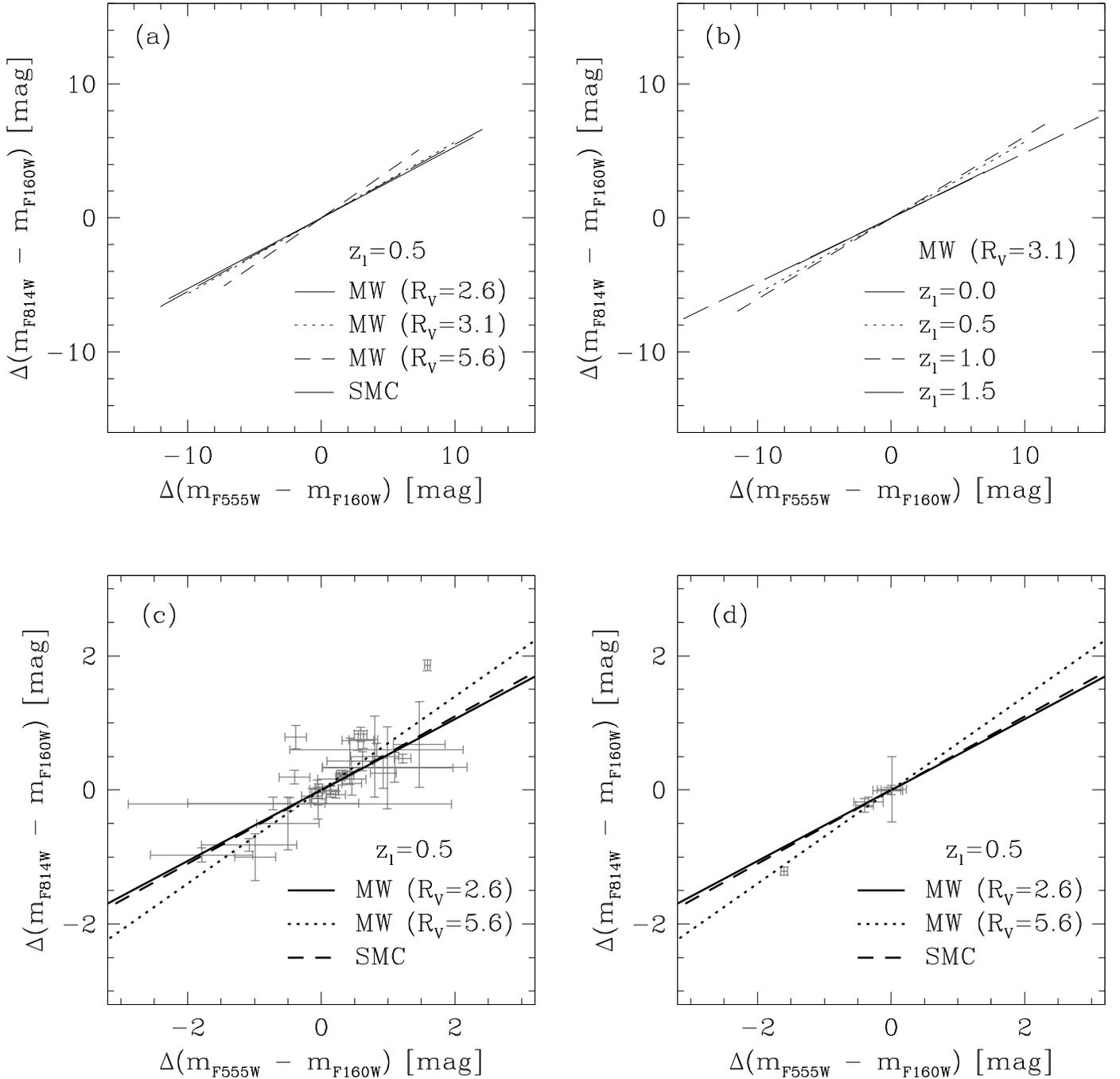}
\caption{The upper two panels show the expected color differences  
between multiple images for various extinction laws 
at the same lens redshift (panel a), 
and for the same extinction law at various lens redshifts (panel b). 
All of these lines are drawn up to the expected color differences 
for $95\%$ of the cumulative differential column density of the gas 
in the lens galaxy, $\Delta N_{\rm H} \sim 7.1 \times 10^{21}~{\rm cm^{-2}}$.
In the upper left panel (panel a), the expected color differences 
for the MW extinction curve with $R_V=2.6$, with
$R_V=3.1$, and with $R_V=5.6$, and for the SMC extinction curve 
are presented by the solid, dotted, dashed, 
and long-dashed lines, respectively.
The applied redshift for the lens in these estimations is $0.5$. 
In the upper right panel (panel b), the expected color difference 
in the case of the lens galaxy at $z_l=0.0$, $0.5$, $1.0$, and $1.5$ 
are presented by the solid, dotted, dashed, 
and long-dashed lines, respectively.
The applied extinction law in this estimation is the MW one with $R_V=3.1$, 
or a standard extinction law in our galaxy.  
The lower two panels show the expected color differences 
for the various extinction laws at the lens redshift of $z_l=0.5$. 
The extinction law for the MW with $R_V=2.6$, the MW with $R_V=5.6$, 
and the SMC are presented by 
the solid, dotted, and dashed lines respectively.  
The observational data which were shown 
in Figure~\ref{fig:color} (a) and (b) is  
overlaid as the gray crosses in panels (c) and (d), respectively.
Note that the scale of the abscissa and ordinate in the upper panels 
is different from that in the lower panels.}
\label{fig:difdust}
\end{figure*}

Denoting the response of the filter $k$ and the spectrum of the source 
as $f_k(\lambda)$ and $S(\lambda)$, respectively, 
the expected magnitude at filter $k$, $m_k$, 
including dust extinction is expressed as  
\begin{equation}
 m_k(N_{\rm H}) = m_{int, k} -2.5 \log \left[ 
  \frac{\int_{\lambda_{\rm min}}^{\lambda_{\rm max}} 10^{-0.4A \left[
  N_{\rm H}, \lambda/(1+z_l) \right] } f_k(\lambda) S(\lambda)
   \, d\lambda}{\int_{\lambda_{\rm min}}^{\lambda_{\rm max}} 
   f_k(\lambda) S(\lambda) \, d\lambda} \right] , 
\label{eq:extinction}
\end{equation}  
where $\lambda_{\rm max}$ and $\lambda_{\rm min}$ represent 
the maximum and minimum wavelengths of the filter response, respectively, 
and $m_{int, k}$ is the source magnitude without dust extinction.  
Finally, the expected magnitude difference between multiple images   
at filter $k$, $\Delta m_k$, due to differential dust extinction 
is expressed as 
\begin{equation}
 \Delta m_k = m_k(N_{\rm H}) - m_k(N_{\rm H}^{\prime}),   
\end{equation}
where $N_{\rm H}$ and $N_{\rm H}^{\prime}$ are the gas column density  
on one image and that on the other image, and 
the difference of these two quantities is $\Delta N_{\rm H}$. 
Consequently, the expected color difference between multiple images 
is estimated by subtracting the magnitude difference of 
one filter from that of the other filter, 
i.e., $\Delta m_k - \Delta m_j$.  
Although the spectral shape of the source is clearly 
involved in equation~\ref{eq:extinction}, dependence on 
the assumed spectral shape causes a difference of only a few percent 
\footnote{By assuming a power-law spectral shape of the source, 
$S(\lambda) \propto \lambda^{- \alpha}$, 
we have checked the dependence. 
The expected dust extinction changes only a few percent 
by changing the power index ($\alpha$) from $0$ to $-3$}. 
Thus, we assume a flat spectrum, 
i.e., $S(\lambda) = \mbox{const}$.  
Applying equation~\ref{eq:extinction} to all the filters,  
we can estimate the expected color differences   
due to differential dust extinction. 
The results are presented in Figure~\ref{fig:difdust}. 

As clearly seen in Figure~\ref{fig:difdust},  
the differential dust extinction of $95\%$ level 
(see Figure~\ref{fig:column}) can produce 
up to $\sim 10~{\rm mag}$ of color difference 
between multiple images of lensed quasars 
\footnote{This value could be too much as a representative value 
for the differential dust extinction, 
but the value indicates the maximum capability of 
the differential dust extinction to produce color differences.}.  
The expected color difference is large enough 
to reproduce the observed chromaticity. 
Moreover, the slope nicely reproduces the observational data. 
We note that the above theoretical predictions
do not depend strongly on the lens redshift and 
the applied extinction laws (see Figure~\ref{fig:difdust}).
However, the expected color difference can be as large as
a few magnitudes, exceeding the observed values for the color difference.
This may be because we have adopted the simulation of a relatively gas-rich
galaxy that has not yet converted a substantial fraction of 
the gas mass into stars \citep[][]{hirashita}. 
Actually, many lens galaxies are suggested to be early type galaxies, 
and the dust amount and distribution in such galaxies must differ from 
what we have applied here. 
In this respect, our approach may be too simplistic.  

In the lower panels of Figure~\ref{fig:difdust},  
some of the expected color differences are presented together with 
the same observational data as those 
in Figure~\ref{fig:color} for comparison. 
Most of the observational data in both of these panels 
(panels c and d are respectively for lensed quasars whose redshift 
 of the lens galaxy is known and unknown) 
are consistent with the the expected color differences  
within the $1\sigma$ error bar. 
This indicates that the observed chromaticity can  
be explained by differential dust extinction.  
However, some data points cannot be reproduced 
by the currently applied extinction laws.   
For example, an upper right data point in panel (c) 
has error bars small enough to conclude that it 
deviates significantly from the theoretical lines. 
We will further discuss such data points in section 6. 

Here, we have applied only well-known empirical formulae 
for the extinction laws, and have not assumed 
any special, eccentric, or other unrealistic extinction laws. 
If it is confirmed that the differential dust extinction is 
responsible for the chromaticity of the lensed quasars, 
and if our assumption that the shape of extinction curve 
is the same at different positions in the lens galaxy, 
is valid \citep[see,][]{mcgough},   
we can conclude that any special extinction laws are 
not required to reproduce dust extinction in galaxies up to $z \sim 1$, 
and the normal extinction laws that we have used in this study, 
i.e., the extinction laws derived from the local Universe, 
can be applicable in such distant galaxies.

\section{Quasar Microlensing}

From macro lensing modeling, the surface mass density of the lens 
divided by the critical surface mass density for lensing 
\citep[e.g.,][]{sef}  
is of order unity for all multiple images of lensed quasars. 
This indicates that the probability of quasar microlensing
is high enough to be observed if the mass is dominated by 
compact components such as stars, planets or other compact objects. 
As investigated by \citet{wam91}, 
quasar microlensing can change the observed color of quasars
from their intrinsic one, and $\mu(t)$ 
in equation~\ref{eq:basic} is also a function of wavelength. 
This effect has been examined by using more realistic models 
for quasar accretion disks \citep[e.g.,][]{yonehara}. 
Since the spatial distribution of lens objects on different images 
is generally different, quasar microlensing occurs 
in different ways in different images and consequently 
the color differences may be produced by quasar microlensing. 
Possible microlensing signals in 
the observed quasar sample are summarized in appendix B. 

Furthermore, the timescale of quasar microlensing can be estimated by 
a timescale on which the source crosses the Einstein ring radius ($t_E$) 
or a timescale on which the caustic crosses the source ($t_c$). 
The former timescale is evaluated by  
\begin{eqnarray}
t_E &=& \frac{r_E}{v_t} \nonumber \\ 
    &=& \frac{1}{v_t} \left( \frac{4GM_{\rm lens}}{c^2} 
          \frac{D_{ls}D_{ol}}{D_{os}} \right)^{1/2} 
\label{eq:ertrans} \\
    &\simeq& 14
             \left( \frac{v_t}{10^3~{\rm km~s^{-1}}} \right)^{-1} 
              \left( \frac{M_{\rm lens}}{1M_{\odot}} \right)^{1/2} 
               \left( \frac{D_{ls}D_{ol}/D_{os}}{1~{\rm Gpc}} \right)^{1/2} 
                {\rm ~yr}, 
\label{eq:ertime}
\end{eqnarray} 

where $r_E$, $v_t$ and $M_{\rm lens}$ represent 
the Einstein ring radius on the lens plane, 
the transverse velocity of the lens on the lens plane, 
and the mass of the lens, respectively \citep[e.g.,][]{irw89}.  
The latter timescale is evaluated by
\begin{eqnarray}
t_c &=& \frac{10^3r_s}{v_c}  \nonumber \\
    &\simeq& 9 \left( \frac{v_c}{10^3 {\rm km~s^{-1}}} \right)^{-1} 
                \left( \frac{M_{\rm BH}}{10^8M_{\odot}} \right)  {\rm ~yr},
\label{eq:cctime}
\end{eqnarray}
where $r_s$, $v_c$ and $M_{\rm BH}$ represent the Schwarzschild radius,  
the transverse velocity of the caustics on the source plane, 
and the mass of supermassive black hole in the center of the quasar,
respectively 
\footnote{The transverse velocity of the caustics is determined 
by a combination of a bulk and a proper motion of the lens objects. 
However, a proper motion of the lens objects dramatically 
changes the caustics networks themselves, 
and the expected transverse velocity is larger than 
the simple combination of these two components \citep[e.g.,][]{jsbw}. 
Moreover, how the caustics networks changes also 
depends on surface density of the lens objects, 
that of smooth matter, and external shear.}.  
Of course, there are several ambiguities in these estimates, 
but the expected timescale for quasar microlensing 
should be comparable to or longer than $1$ year. 
This timescale is long enough for the color difference 
caused by microlensing to be observed as a static phenomenon, 
and the expected color difference due to quasar microlensing 
has to be a realistic candidate to reproduce the observed chromaticity. 

To quantify the microlensing effect, 
we should treat the continuum source (quasar accretion disk) and 
the magnification properties of microlensing. 
For a model of the continuum source, we applied the so-called 
``standard accretion disk model \citep[][]{shsu}'' 
as a central engine of quasars. 
Here, the inner and the outer radii of the accretion disk 
are set to be $3r_s$ and $10^3r_s$ 
\footnote{The effective temperature of the accretion disk at
a radius of $10^3r_s$ is $\sim 2600~{\rm K}$, 
and the peak wavelength of black body spectrum 
with this temperature is $\sim 1.1 \times 10^4~\AA$. 
Assuming the source redshift to be $z_s=1.0$, 
the peak wavelength at observer is 
$\sim 1.1 \times 10^4 \times (1+z_s) = 2.2 \times 10^4~\AA$. 
Since this is longer than wavelength coverage of the reddest filter 
in our calculations (F160W of {\it HST} NICMOS), 
the outer radius larger than $\sim 10^3r_s$ does not change the
results. If the radius is much smaller than $10^3r_s$, the
resulting microlensing signal tends to be enhanced. }, respectively, 
and the accretion rate is set to be a critical value.   
Thus, the only parameter specifying the properties of an accretion disk 
is the mass of the central supermassive black hole ($M_{\rm BH}$).   
Some emission lines can contribute to the optical flux, 
and such emission lines can be affected by microlensing in some cases
\citep[e.g.,][]{abajas,lewis,richards,sluse}. 
However, it is hard to imagine that all the sample
quasars with various redshifts are significantly affected
by line emissions in the observational bands. Here,
we neglect such line emissions and consider only continuum
emissions. 
For magnification properties, we have considered 
the properties of the caustics, 
which corresponds to regions where the source is 
extremely magnified by microlensing. 
In this paper, we take a straight line approximation 
for fold caustics and apply an approximated magnification 
in the vicinity of caustics which is only a function of 
a distance from the caustics \citep[e.g.,][]{sef}.  
Including a constant magnification,  
(magnification due to all the lensing effects except quasar microlensing), 
the magnification applied in this study is expressed as 
\begin{eqnarray}
\mu(x) &=& \left( \frac{x}{x_s} \right)^{-1/2} + \mu_a 
 {\rm ~ ~ ~ (inside~ ,or~positive~side~of,~the~caustics)} 
\label{eq:qmlappmag} \\
       &=& \mu_a {\rm ~ ~ ~ (outside~ ,or~negative~side~of,~the~caustics)},
\label{eq:appcaus}
\end{eqnarray}  
where $x$, $x_s$, and $\mu_a$ represent the 
distance from the caustics, the scale length of the caustics  
and the constant magnification, respectively
\footnote{When the source crosses fold caustics and comes 
`inside' the caustics, 
two images appear at corresponding critical curves. 
If the source moves to the other way around, 
i.e., the source goes `outside' the caustics, 
the pair of images will disappear at the critical curves.   
Even a part of the caustics is approximated as a straight line, 
such property should remain unchanged and  
is involved in equation~\ref{eq:appcaus}.  
The first term in the right hand side of equation~\ref{eq:qmlappmag} 
represents magnification for the appearing/disappearing pair of images 
at the critical curves.}.   
Since this formula is investigated from general 
(mathematical) properties of gravitational lensing, 
this approximation is applicable not only for 
caustics produced by single lens object, but also for more general case 
such as caustics of quasar microlensing, 
i.e., caustics produced by multiple lens objects 
with external convergence and shear. 
Effects of the distribution of lens objects, external convergence 
and shear on magnification are hidden in 
the scale length, $x_s$, in this approximation  
 (see more details in Appendix C).   

\begin{figure}
\centering
\includegraphics[width=\hsize]{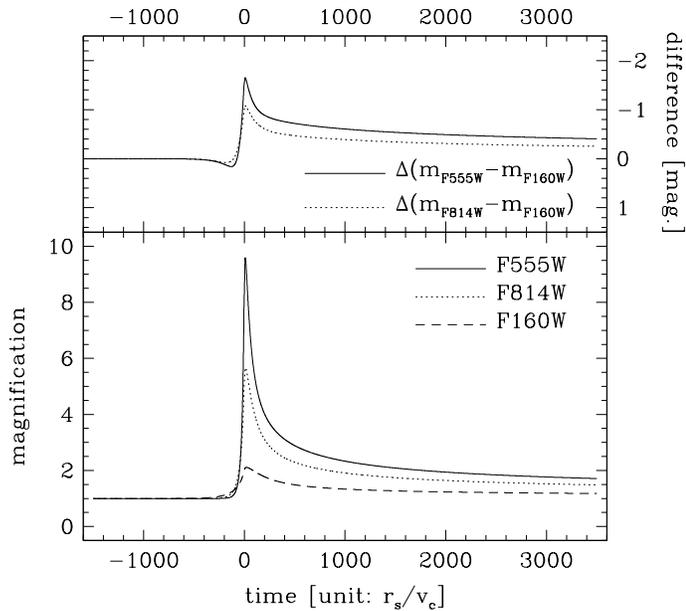}
\caption{In the lower panel, the expected quasar microlensing 
light curves at F555W, F814W, and F160W are presented by 
the solid, dotted, and dashed line, respectively. 
Abscissa is time steps in unit of $r_s/v_c$, and 
ordinate is magnification factor in linear scale.  
At the beginning of these light curves, 
the source is completely outside the assumed caustics 
and the location of the source center respective to the caustics is 
$-1.5 \times 10^3 r_s$. 
At the end of these light curves, 
the source is completely inside the assumed caustics 
and the location of the source center respective to the caustics is 
$+3.5 \times 10^3 r_s$. 
In the upper panel, corresponding color changes (in mag)
during the events are also presented by 
the solid (F555W and F160W) and dotted (F814W and F160W) line. 
$z_s$, $M_{\rm BH}$ and $\mu_a$ are set to be $2$, $10^8M_{\odot}$, and $1$, 
respectively.
The Einstein ring radius for $z_l=0.5$ and for $M_{\rm lens}=1M_{\odot}$  
at the source plane is applied for $x_s$.}  
\label{fig:qmldemo}
\end{figure}

Magnification due to macrolensing is different at different images, 
and $\mu_a$ is also different at different images. 
Then, the expected magnitude at filter $k$ at epoch $t$, $m_k$, 
including quasar microlensing magnification is expressed as
\begin{eqnarray}
 &m_k(t, \mu_a) =& \nonumber \\ 
 &m_{int, k} - 2.5& \log \left[ 
 \frac{\int_{\lambda_{\rm min}}^{\lambda_{\rm max}} f_k(\lambda)
 \int_{\rm area} \mu (|{\bf x} - {\bf x_c}(t)|_{\rm min}) 
  \epsilon_{\lambda/(1+z_s)} ({\bf x}) \, d{\bf x} \, d\lambda}
   {\int_{\lambda_{\rm min}}^{\lambda_{\rm max}} f_k(\lambda) \int_{\rm area} 
    \mu_a \epsilon_{\lambda/(1+z_s)} ({\bf x}) \, d{\bf x} \, d\lambda} 
\right], 
\label{eq:qmlmagc}
\end{eqnarray}
where $\epsilon_{\lambda}({\bf x})$ and ${\bf x_c}(t)$ represent 
emissivity distribution of the accretion disk at wavelength $\lambda$ 
and the 2-dimensional location of caustics on the sky 
at an observational epoch, respectively. 
Finally, the expected magnitude difference 
between multiple images at filter $k$, $\Delta m_k$, 
due to quasar microlensing is expressed as 
\begin{equation}
 \Delta m_k = m_k(t, \mu_a) - m_k(t^{\prime}, \mu_a^{\prime}), 
\end{equation}
where $t$ and $t^{\prime}$ are an epoch on microlensing at one image 
and that at the other image, respectively, and 
they should be measured at the rest-frame of lens galaxy. 
Again, the expected color difference between multiple images 
are estimated by subtracting the magnitude difference 
at one filter from the other filter, i.e., $\Delta m_k - \Delta m_j$. 
$\mu_a$ and $\mu_a^{\prime}$ are a macrolensing magnification factor of 
one image and that of another image, respectively. 
The inner integral should be performed all over the source, 
and the term $|{\bf x} - {\bf x_c}(t)|_{\rm min}$,  
minimum value for $|{\bf x} - {\bf x_c}(t)|$, 
corresponds to the distance to the caustics. 
In this estimate, like the estimation for differential dust extinction, 
we again take into account the filter responses.
An example of microlensing light curves for all the three filters 
are presented in Figure~\ref{fig:qmldemo}. 
Since there is no correlation among location of caustics on different images, 
${\bf x_c}$ must be different at different images. 

Based on the above definitions and assumptions, 
we have calculated microlensing light curves 
in all the three filters for various parameters. 
The light curves show a clear waveband dependence 
in their shapes (see Figure~\ref{fig:qmldemo});
at shorter wavelength, the expected magnification changes more dramatically 
and the maximum magnification becomes larger.  
This is because the emission at a shorter wavelength 
comes only from a relatively inner compact region of the accretion disk 
compared to the emission at a longer wavelength 
\citep[e.g.,][]{shsu}.    
A smaller source is magnified more in a microlensing event 
because a large fraction of the source can locate inside 
a strongly magnified region. 
Further, as shown in equation~\ref{eq:cctime}, 
the timescale of microlensing event depends on the source size, 
and the event observed at shorter wavelength is rapid 
\citep[][]{wam91, yonehara}. 
For all calculations of the light curves, 
the source center is located at $-1.5\times10^3r_s$ 
from the caustics at the beginning 
and is located at $+3.5\times10^3r_s$ 
from the caustics at the end. 
Since we assume $10^3r_s$ as the accretion disk size, 
the source is completely outside the caustics at the beginning, 
and is completely inside the caustics at the end.

\begin{figure*}
\centering
\includegraphics[width=\textwidth]{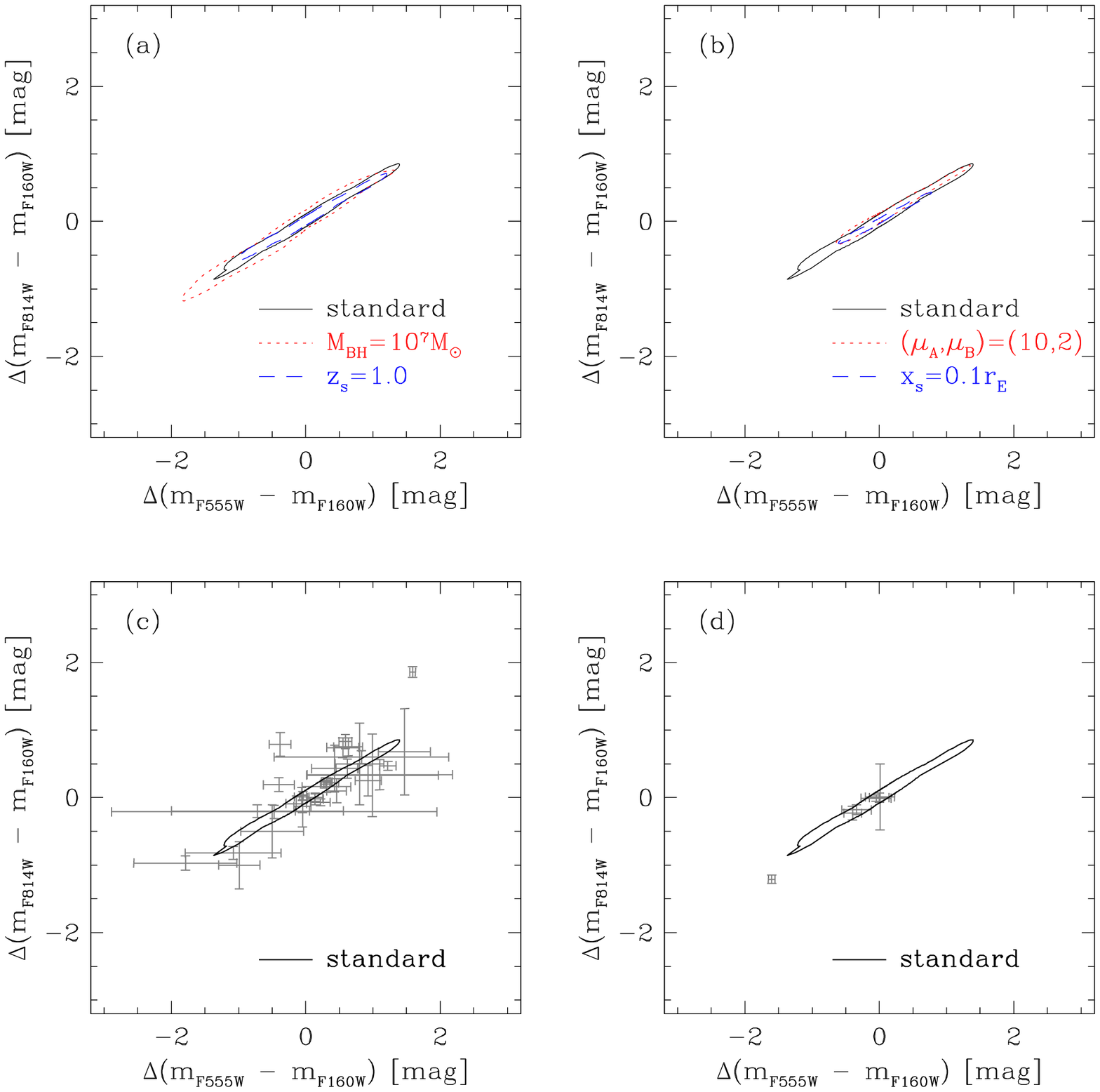}
\caption{The expected color differences between multiple images   
due to quasar microlensing. 
The maximum color differences by quasar microlensing are 
presented by the lines. 
The upper two panels show the expected color difference for various 
parameter sets; $M_{\rm BH}$, $z_s$, $\mu_A$, $\mu_B$, and $x_s$.
The solid line in all panel represents the expected color difference for 
$M_{\rm BH}=10^8M_{\odot}$, $z_s=2.0$, $(\mu_A, \mu_B)=(3.0, 2.0)$, 
and $x_s=r_E$, and is referred to as `standard' in all the panels.  
$M_{\rm lens}=1M_{\odot}$ and $z_l=0.5$ are applied 
for calculating $r_E$ in all the cases, and $r_E$ in this estimation 
is not on the lens plane, but on the source plane. 
In the upper left panel, panel (a), the expected color differences 
for $M_{\rm BH}=10^7M_{\odot}$ and for $z_s=1.0$ 
are presented by the dotted (red) and dashed (blue) lines, respectively. 
In the upper right panel, panel (b), 
the expected color difference for $(\mu_A, \mu_B)=(10,2)$ and for $x_s=0.1r_E$ 
are presented by the dotted (red) and the dashed (blue) line, respectively. 
The remaining parameters are the same as that 
for parameters of the `standard' case.   
Observational data shown in Figure~\ref{fig:color} (a) and (b) 
is plotted by gray crosses in panels (c) and (d), respectively, 
and the expected color differences for $M_{\rm BH}=10^8M_{\odot}$ presented 
in panel (a) are overlaid on these panels by the solid line.}
\label{fig:qmlcolor}
\end{figure*}

The range completely covers the most interesting epoch 
for quasar microlensing events associated with 
caustic crossing, 
and the coverage is enough for our current purpose. 
Subsequently, we pick up all possible combinations of 
two epochs in the light curve. 
This mimics an observation of two images in multiple quasars 
with different time delay. 
Subtracting the magnitudes of the brighter image at F160W 
from that of the fainter one in all filters, 
and calculating the color difference between the images, 
we can estimate how much color difference between images 
is expected from quasar microlensing. 
The expected color difference due to quasar microlensing 
in our calculations is determined by 5 parameters; 
$M_{\rm BH}$, $z_s$, $\mu_a$ for two images to be compared 
(denoted as $\mu_A$ and $\mu_B$), and $x_s$. 
Based on realistic estimations, 
$x_s$ takes values around $r_E$ and  
we apply $r_E$ as a standard value of $x_s$ 
(see Appendix C for more details about $x_s$).   
Since $r_E$ is a function of $M_{\rm lens}$, $z_l$ and $z_s$, 
physical value of $x_s$ is determined by $M_{\rm lens}$, $z_l$ and $z_s$. 
The expected area of the color differences are shown 
in Figure~\ref{fig:qmlcolor}.  

It is apparent from Figure~\ref{fig:qmlcolor} that 
the maximum color difference depends strongly on  
$M_{\rm BH}$, $(\mu_A, \mu_B)$ and $x_s$, and weakly on $z_s$. 
Within a reasonable range for $z_s$ (e.g., see Figure~\ref{fig:redshift}), 
the rest frame wavelength which corresponds to 
the observational waveband varies by a factor of few. 
Thus, emissivity distributions for the observational wavebands 
at the different source redshift differs slightly.  
Not only due to the smaller source size, 
but also due to the bluer spectrum of the accretion, 
the expected color change for smaller $M_{\rm BH}$ 
shows different features than that for larger $M_{\rm BH}$, 
as shown by the dotted line in Figure~\ref{fig:qmlcolor} (a). 
For larger $\mu_a$, i.e., larger $\mu_A$, larger $\mu_B$ or both 
in these calculations, the color difference is suppressed 
as shown by the dotted line in Figure~\ref{fig:qmlcolor} (b). 
This can be understood from equations~\ref{eq:qmlappmag},
~\ref{eq:appcaus} and~\ref{eq:qmlmagc}. 
Magnitude changes due to quasar microlensing are determined by 
a ratio of a magnification for inside the caustics 
(equation~\ref{eq:qmlappmag}) 
to a magnification for outside the caustics 
(equation~\ref{eq:appcaus})  
rather than a magnification for inside the caustics alone 
(see equation~\ref{eq:qmlmagc}).  
Since the contribution of the term including $x$ 
in equation~\ref{eq:qmlappmag} 
to the magnitude change is relatively smaller in the larger $\mu_a$ case, 
the change due to microlensing becomes small, 
and consequently, the color change is smaller 
compared to the smaller $\mu_a$ case.  
This means that the magnification due to macrolensing 
could be crucial to quantify 
how much color difference would be expected from quasar microlensing. 
Of course, another major factor to estimate the color difference 
is a property of magnification pattern 
as shown with the dashed lines in Figure~\ref{fig:qmlcolor} (b). 
If we apply smaller $x_s$, only a part of the source is   
magnified by microlensing, and a smaller amount of 
color difference is expected (see equation~\ref{eq:appcaus}). 

Since the `standard' parameter set in Figures~\ref{fig:qmlcolor} (a) and (b) 
can be a representative parameter set for actual quasar microlensing events, 
the expected color is also plotted with observational data 
in Figures~\ref{fig:qmlcolor} (c) and (d) for comparison. 
Quasar microlensing alone can also produce 
color differences up to $\sim 2{\rm ~mag}$ 
which is similar to the scatter of observational data. 
Further, the observed color difference is well reproduced 
by quasar microlensing within the error bars except 
some individual data points, for instance 
the upper right data point in Figure~\ref{fig:qmlcolor} (c) 
and the lower left data point in in Figure~\ref{fig:qmlcolor} (d). 

\begin{figure*}
\centering
\includegraphics[width=\textwidth]{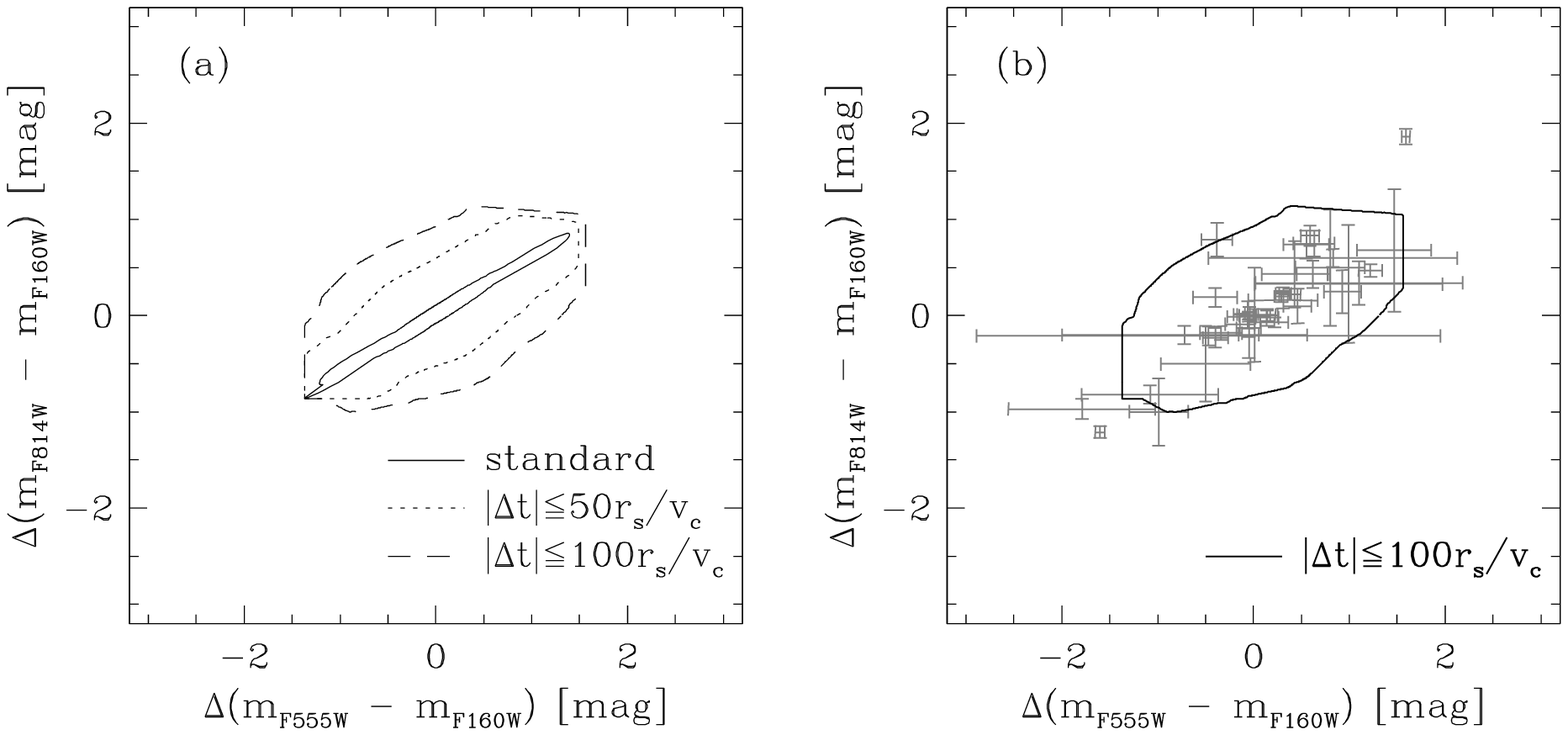}
\caption{The expected color differences between multiple images  
due to quasar microlensing.  
Here, possible ranges of observations are included. 
`Standard' set of parameters in Figure~\ref{fig:qmlcolor} 
are also applied for these calculations, 
and the result obtained without including the time difference, 
i.e., $|\Delta t| = 0$, is presented as a reference 
by the solid line in the left panel, panel (a). 
In panel (a), the result for $|\Delta t| < 50 r_s/v_c$ and 
$|\Delta t| < 100r_s/v_c$ is also presented by 
the dotted and the dashed line, respectively. 
Observational data shown in Figure~\ref{fig:color} (a) and (b) 
are plotted by gray crosses in panels (b).  
The expected color difference in the case of $|\Delta t| < 100r_s/v_c$ 
is also overlaid on this panel by the solid line.}
\label{fig:qmltadd}
\end{figure*}

As is already noted in the case of quasar variabilities (section 3), 
the date of observations is different in different wavebands   
\citep[e.g.,][]{lehar}, and we should take into account 
the interval among the multi-waveband observations for one object, $\Delta t$.
However, conversion of the timescale of microlensing 
used in numerical calculation, e.g., see equation~\ref{eq:cctime}, 
into an actual timescale involves some ambiguities.  
Here, we applied $|\Delta t| \le 50 r_s/v_c$ and $|\Delta t| \le 100 r_s/v_c$ 
as the time difference with respect to the date of observation at F555W  
and calculate the expected color difference between images.  
That is, denoting the observational epoch at F555W  
in the microlensing light curve for two images by $t_{A}$ and $t_{B}$, 
the observational epoch for two images at F814W and F160W is 
$t_{A} + \Delta t$ and $t_{B} + \Delta t$, and 
$t_{A} + \Delta t^{\prime}$ and $t_{B} + \Delta t^{\prime}$,
respectively, where  
$\Delta t$ and $\Delta t^{\prime}$ are uncorrelated and 
all possible combinations for $\Delta t$ and $\Delta t^{\prime}$ 
within the applied time difference are chosen for calculations.   
The maximum time difference of $|\Delta t| \le 50 r_s/v_c$ 
and $|\Delta t| \le 100 r_s/v_c$ 
corresponds to $\sim 1~{\rm yr}$ and $\sim 2~{\rm yr}$, respectively, 
in the case of $v_c=10^3~{\rm km~s^{-1}}$ and $M_{\rm BH}=10^8M_{\odot}$.   
The results are presented in Figure~\ref{fig:qmltadd}.  
As easily seen in Figure~\ref{fig:qmltadd}, 
the expected color difference with non-zero time difference among  
different wavebands (the dotted and the dashed lines 
in Figure~\ref{fig:qmltadd}a) is stretched toward an orthogonal direction 
to the major axis of the expected color difference calculated
without time difference (the solid line in Figure~\ref{fig:qmltadd}a). 
An effect of time difference largely extends 
the theoretically allowed region of the expected color difference 
produced by quasar microlensing. 
Consequently, most of the observed color difference of multiple quasars 
including objects whose observed color difference cannot be reproduced 
by differential dust extinction are also nicely reproduced 
(Figure~\ref{fig:qmltadd} (b)) 
\footnote{The expected color differences presented here 
are the maximum color differences which 
can be produced within the framework of our calculations. 
Since we do not include any statistical properties of caustics 
such as distribution of the scale length of the caustics    
or clustering of the caustics, 
the expected color differences can not reproduce  
distribution of the data points well.  
This could be the reason why the data points are clustering 
rather narrow region in Figure~\ref{fig:qmltadd} (b) 
compared to the expected color differences.}.

Since we do not include any statistical properties of caustics 
such as distribution of the scale length of the caustics      
or clustering of the caustics 
the expected color differences can not reproduce  
distribution of the data points well 
\footnote{This could be the reason why the data points are clustering 
rather narrow region in Figure~\ref{fig:qmltadd} (b) 
compared to the expected color differences.}. 
On one hand, therefore, we cannot directly compare to 
the results for quasar microlensing to the distribution of 
observational data 
(e.g., data points shown in Figure~\ref{fig:qmltadd} (b)).  
The expected color differences presented here 
(e.g., a closed curve shown in Figure~\ref{fig:qmltadd} (b)) 
are the maximum color differences which can be produced within 
the framework of our calculations with a given $x_s$.   

On the other hand, the applied approximation for magnification 
due to quasar microlensing is surprisingly sufficient 
to reproduce the observed color differences. 
Our treatment, an approximation for magnification based on general 
properties of gravitational lensing, is rather simple 
compared to more realistic methods such as ray-shooting, 
and does not include any fine structure of the caustics 
such as curvature of fold caustics, cusp caustics
\footnote{Since cusp caustics require more special conditions
than fold caustics (Schneider et al.\ 1992), 
fold caustics crossing occurs much more frequently  
compared to cusp caustics crossing. 
Thus, ignoring cusp caustics crossing only misses
some rare events. }, 
or an overlap of caustics.  
We suspect that our simple treatment is enough to 
reproduce the observed color differences since
the observed color differences 
are mainly produced by isolated caustics and/or 
caustics with larger size compared to fine structures of caustics.
Even if caustics with a typical size smaller than the source size
overlap with caustics with a larger size, the former (smaller) 
caustics add only a tiny fluctuation of magnification, 
and cannot produce significant amount of color differences.   
Thus, we can interpret that our simple approach
extract only the caustics which contribute to the strongest
magnification at a certain moment. 
This putative explanation should be checked 
by more sophisticated calculations in future. 

Even with quasar microlensing, it is clear that  
the observed color differences of some extreme objects 
which are located at a top-right corner and a bottom-left corner 
in Figure~\ref{fig:qmltadd} (b) are not well reproduced.   
We will further discuss this issue in section 6.

\section{Discussion}

In this paper, we have examined the following three possibilities
for explaining the observed chromaticity  
between multiple images of lensed quasars: 
(i) intrinsic variabilities of the lensed quasars, 
(ii) differential dust extinction in the lens galaxy, 
and (iii) quasar microlensing in the lens galaxy.

\subsection{Overall trends}

First, we have examined the intrinsic variabilities of the lensed quasars. 
Quasars generally have intrinsic variabilities, 
and there exists a time delay between multiple images of lensed quasars. 
Thus, the intrinsic variabilities of quasars could explain
the observed color differences. 
However, we find that the expected value of the color difference is 
roughly one order of magnitude smaller than the observed color differences. 
Therefore, we can exclude the possibility that intrinsic variabilities are 
responsible for the observed color differences in the sample.  

\begin{figure*}
\centering
\includegraphics[width=\textwidth]{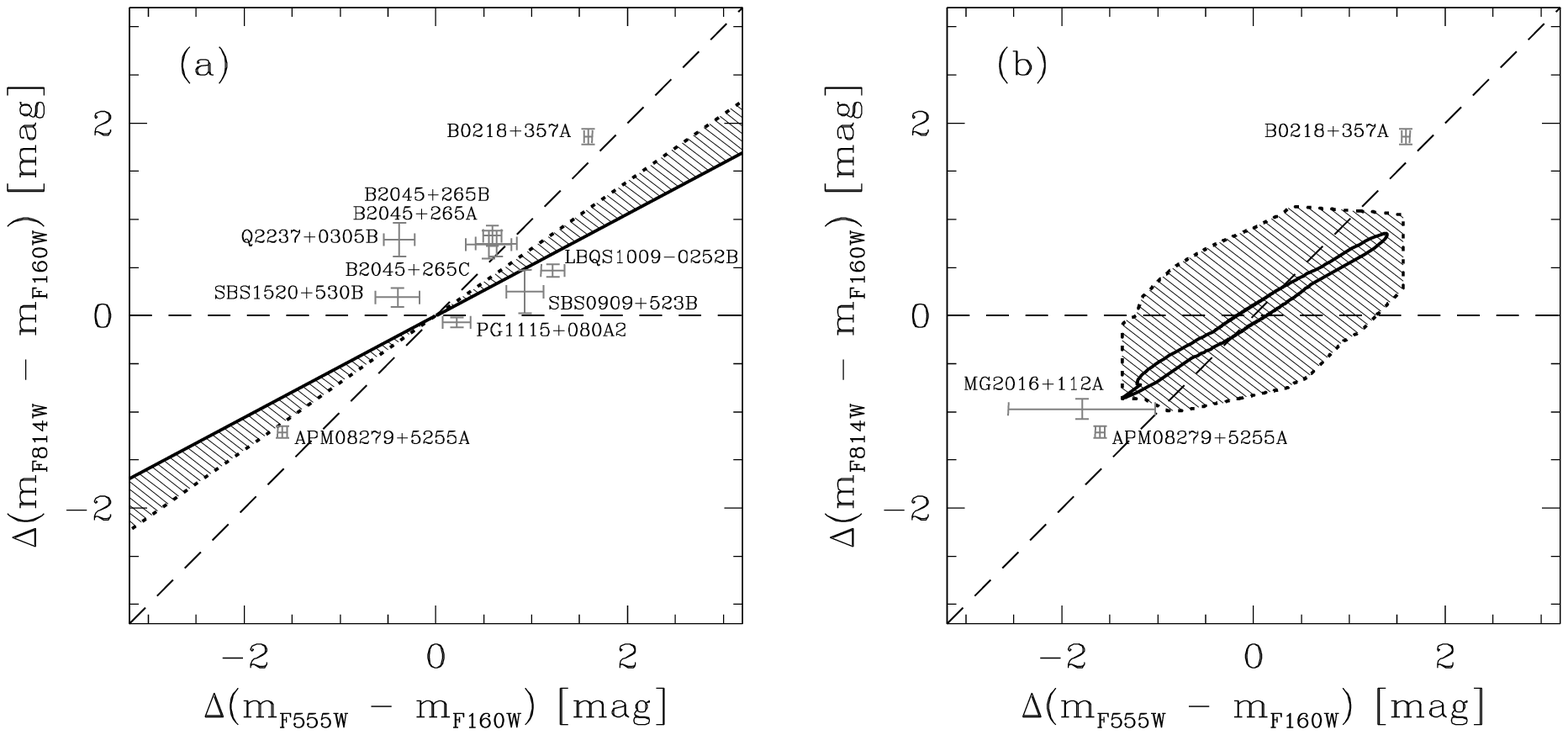}
\caption{Objects whose observed color difference 
between multiple images is not
well reproduced by our estimations are picked out and 
are presented with their name, and 
with the expected color difference by our estimations.  
In the left panel (panel a), 
the expected color difference for the MW extinction curve with $R_V=2.6$ 
and that for MW with $R_V=5.6$ are presented by 
the solid and dotted lines, respectively, where $z_l=0.5$ is assumed. 
The expected color difference locates 
between the solid and the dotted line, i.e., the shaded region. 
Objects whose observed color difference cannot be reproduced
by differential dust extinction are also presented by the gray crosses. 
In the right panel (panel b), 
the expected color difference for `standard' case of quasar microlensing 
and that including difference in observational epochs at different wavebands 
are presented by the solid and dotted lines, respectively, 
where $z_s=2.0$ is assumed. 
The expected color difference for `standard' case with the non-zero spans 
locates inside the solid line, i.e., the shaded region. 
Again, objects whose observed color difference cannot be reproduced
by quasar microlensing are presented by the gray crosses. 
The line for $\Delta m_{F814W} - \Delta m_{F160W}=0$ and the line for 
$\Delta m_{F555W} - \Delta m_{F160W} = \Delta m_{F814W} - \Delta m_{F160W}$ 
are also presented by the dashed lines as references.}
\label{fig:discuss}
\end{figure*}

As shown in Figures~\ref{fig:difdust} and~\ref{fig:qmlcolor}, 
both differential dust extinction and quasar microlensing 
instead can reproduce the observed color differences. 
The color differences due to these two scenarios  
similarly reproduce the observational data points 
in Figures~\ref{fig:difdust} and~\ref{fig:qmlcolor}.  
However, once we take into account the time difference among the observations 
at different wavebands that usually happens in an actual situation, 
the area covered by the expected color difference of quasar
microlensing expands, and we can reproduce the observed color difference 
by quasar microlensing alone, as shown in Figure~\ref{fig:qmltadd}.  
These results are also summarized in Figure~\ref{fig:discuss}; 
objects whose observed color differences are not reproduced 
by a single scenario alone, either differential dust extinction 
or quasar microlensing, are plotted with the expected color difference 
by our estimations.  
Although there is an ambiguity in converting a timescale used 
in our numerical calculations into an actual timescale, 
quasar microlensing proves to be a better scenario 
to reproduce observed color differences compared to other scenarios,  
if only one scenario is responsible for the observed color differences. 
In an actual situation, it is more probable that 
both effects 
play a role at the same time in producing the observed color differences. 
Combining these two scenarios (differential dust extinction and 
quasar microlensing), all the observed color differences 
presented in this paper seem to be nicely reproduced 
(see Figure~\ref{fig:discuss}); 
if we convolve the shaded region in Figures~\ref{fig:discuss} (a) and (b), 
we may be able to reproduce the observed color differences (the gray crosses)  
shown in Figure~\ref{fig:discuss}  
which cannot be reproduced by either differential dust extinction or 
quasar microlensing.
 
Unfortunately, it is not possible to discriminate which one is 
dominating the observed color differences with   
only one photometric observation as used in this paper. 
Since differential extinction is a static phenomenon  
and quasar microlensing is a temporal phenomenon whose timescale is
given by equations~\ref{eq:ertime} and/or~\ref{eq:cctime}, 
multiple or monitoring observations of lensed quasars 
at multiple wavebands are the only way to distinguish between  
differential dust extinction and quasar microlensing. 
Multi-wavelength data taken at the same time 
or during negligibly short spans of observations compared 
to timescale of quasar microlensing are also required 
to suppress the color diversity caused by the time difference
among the observations at different wavebands. 

Of special interest are the properties of objects
whose observed color differences are not reproduced by 
either differential dust extinction or quasar microlensing. 
Even if the observed color difference is within 
the expected color difference of our estimations, 
we have to care about quasar microlensing and/or 
intrinsic variabilities of quasars. 
Thus one should trace and separate effects of quasar microlensing 
and/or intrinsic quasar variabilities properly 
to derive the extinction curve of the lens galaxy: 
otherwise the extinction curves derived for lens galaxies may be 
contaminated by the microlensing and/or 
the intrinsic quasar variability effects, 
and it will be an unrealistic at the worst.

\subsection{Peculiar features} 

Next, we focus on objects whose observed color differences 
are difficult to be reproduced by either differential dust extinction 
or quasar microlensing.  
In Figures~\ref{fig:discuss} (a) and (b), 
we choose 8 objects (10 image pairs) whose color differences 
are not explained by differential dust extinction alone   
(B0218+357, SBS0909+523, LBQS1009-0252, PG1115+080, 
SBS1520+530, B2045+265, Q2237+0305, and APM08279+5255), 
and 3 objects (3 image pairs) whose color differences 
are not explained by quasar microlensing alone  
(B0218+357, MG2016+112, and APM08279+5255).

\begin{figure*}
\centering
\includegraphics[width=\textwidth]{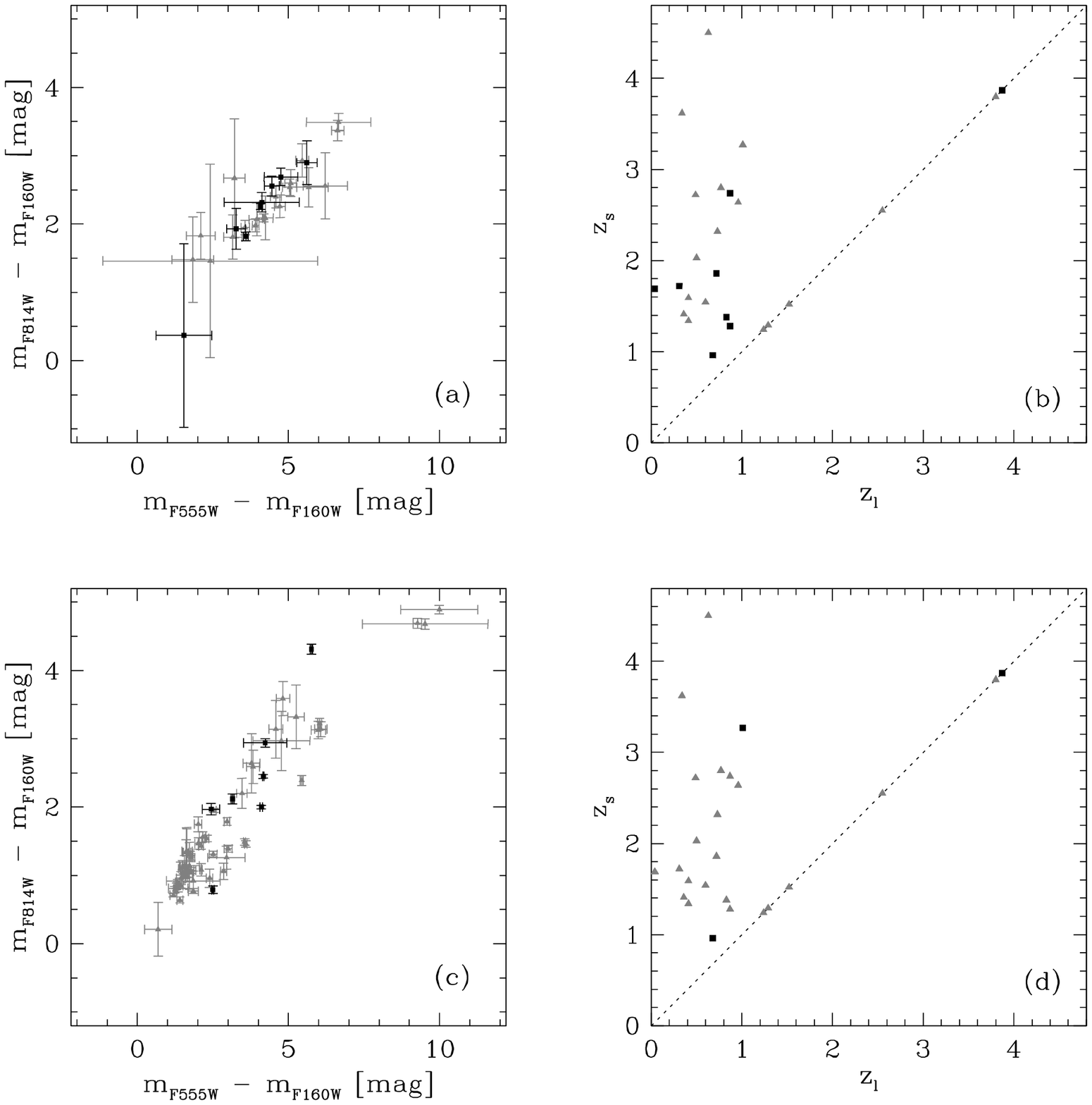}
\caption{Relations between the quantities concerning the lens
galaxies and the source quasars.  
The black filled squares in the upper and lower panels represent 
the objects that have deviation from the expected color difference 
by differential dust extinction and by quasar microlensing, respectively.   
Properties for the remaining objects are also 
plotted by the gray filled triangles.  
The color-color diagram for the lens galaxies and the source quasars 
are presented in panels (a) and (c), respectively.
In panels (b) and (d), the lens redshift and the source redshift 
are presented in the same way as panels (a) and (c).  
The objects with unknown lens redshifts 
are plotted on the diagonal dotted line.}  
\label{fig:prop}
\end{figure*}

A possible explanation for eight objects deviating from the
expected color difference in the differential dust extinction scenario    
is that the lens galaxies have a peculiar dust extinction which cannot be 
parameterized by the function that we have used in this paper.  
Since galaxies at different evolutionary stages  
are expected to have dust with different extinction properties
\citep[e.g.,][]{maiolino04, hirashita05},  
this would be one plausible explanation. 
Furthermore, \citet{inoue} have shown that the wavelength 
dependence of the dust attenuation is modified by a effect of 
light scattering.  
An anomalous dust extinction (or attenuation) might be seen in 
a systematic difference in the colors of lens galaxies themselves. 
Thus, in order to check if the anomalous dust
extinction is really responsible for the deviation from the model
predictions, we examine the colors of the lens
galaxies in Figure~\ref{fig:prop} (see also Table~\ref{tab:gal}). 
Photometric data for the lens galaxies of B1600+434 and Q1208+101 are  
available not for all the bands, and the data are not plotted in
Figure~\ref{fig:prop}.   
B1030+071, MG2016+112, FBQ0951+2635, and Q0957+561 have 
another lens candidate, and the data are also plotted 
in Figure~\ref{fig:prop}. 

Unfortunately, as we can see in Figure~\ref{fig:prop}(a), 
this explanation does not seem to be confirmed.  
Colors of the lens galaxy of objects 
with the unexpected color differences
and that with the expected color differences 
are compared in Figure~\ref{fig:prop}(a). 
The average color for the former objects are 
$(m_{F555W}-m_{F160W}, m_{F814W}-m_{F160W})=(3.93\pm1.12, 2.11\pm0.74)$, 
and that for the latter objects are 
$(m_{F555W}-m_{F160W}, m_{F814W}-m_{F160W})=(4.35\pm1.40, 2.33\pm0.53)$.  
The average color of the lens galaxy of objects 
with the unexpected color differences 
is slightly bluer than the others, 
but these values are overlapping each other within the dispersion.  
Thus, we can say that these two groups are not separated clearly
as far as the galaxy color is concerned. 
Further, there is no clear difference between 
the redshifts of these two groups. 
The average redshifts are $(z_l, z_s)=(0.62\pm0.30, 1.94 \pm 0.88)$ and 
$(z_l, z_s)=(0.61\pm0.23, 2.36 \pm 0.97)$ for 
the objects with the unexpected color differences 
and the others, respectively (see also Figure~\ref{fig:prop} b).

In principle, dust extinction should be larger at shorter wavelength. 
This limit corresponds to the dashed lines in Figure~\ref{fig:discuss}; 
the horizontal line indicates that absorption at $F814W$ and 
that at $F160W$ is equal, and the diagonal line shows
that absorption at $F555W$ and that at $F814W$ are equal. 
Therefore, the color differences by dust extinction should always be 
between the dashed lines in Figure~\ref{fig:discuss}(a).  
In other words, the color difference of objects that are not located 
in this region is not reproduced by dust extinction alone.  
As shown in Table~\ref{tab:gal}, 
some of the lens galaxies are known to have 
observational properties of late-type galaxies; 
B0218+357, RXJ0911+0551, B1030+071, SBS1520+530, 
B1600+434, B2045+265, and MG2016+112, 
and dust properties in such systems may be  
somehow different from that in the MW or the SMC. 
It is still possible to explain the color differences in B2045+265 
by differential extinction of dust with special properties  
because the color differences are located at ``allowed'' area  
for differential dust extinction. 
However, color differences in B0218+357, SBS1520+530, and Q2237+0305 
are located at ``prohibited'' area for differential dust extinction, 
and their anomalous color differences are hard to be explained 
by differential dust extinction alone, 
if dust extinction is assumed to cause reddening in any wavelength. 
However, as can be seen from Figure \ref{fig:prop}, 
there is no clear evidence for peculiarity in the colors and 
redshifts of the objects whose deviation from the models is 
large with respect to other objects. 

As for the 3 objects deviated from the expected color difference 
in the quasar microlensing scenario, it is possible that
the lensed quasar have a peculiar physical properties of 
an accretion disk which cannot be well represented by 
the standard accretion disk model that we have used in this paper.  
Since the expected color differences by quasar microlensing 
may depend on the radiative properties of the central engine of quasars, 
a systematic difference in the source color between these two groups 
would be expected. 
In Figure~\ref{fig:prop}(c), we present the color-color diagram of 
all the source quasar images.  
Again, however, no clear difference between these two groups is found. 
The average colors of the source quasars of the systems
with unexpected and expected color differences from 
our microlensing models are 
$(m_{F555W}-m_{F160W}, m_{F814W}-m_{F160W})=(3.76\pm1.08, 2.37\pm1.00)$  
and $(m_{F555W}-m_{F160W}, m_{F814W}-m_{F160W})=(3.03\pm2.31, 1.75\pm1.16)$, 
respectively. 
Thus, as long as we focus on the colors of the source quasars, 
we cannot find any significant difference between these two groups. 
Further, there is no clear difference between 
the redshifts of these two groups. 
The average redshifts are $(z_l, z_s)=(0.85\pm0.17, 2.70 \pm 1.25)$ and 
$(z_l, z_s)=(0.58\pm0.24, 2.16 \pm 0.90)$ for 
the objects with the unexpected color differences 
and the others, respectively (see also Figure~\ref{fig:prop} d).

\begin{figure}
\centering
\includegraphics[width=\hsize]{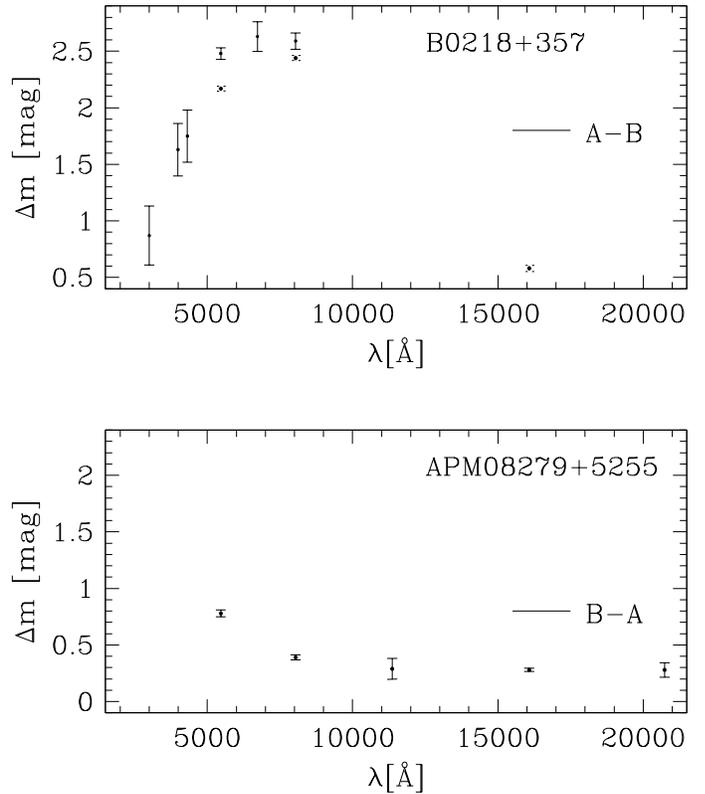}
\caption{Magnitude difference at several wavebands between 
images of B0218+357 (upper panel) and APM08279+5255 (lower panel). 
These data are also provided by CASTLES web page. 
For B0218+357, magnitude differences (magnitude of image A minus 
that of image B, the brightest image at F160W of {\it HST} NICMOS) 
at F300W, F380W, F439W, F555W, F675W, and F814W of {\it HST} WFPC2 
are presented by the solid error bars. 
Data used previously in this paper are different sequence from 
these photometric data at 6 wavebands, and the data used previously 
are also presented by the dashed error bars as references. 
For APM08279+5255, magnitude differences at F555W and F814W of 
{\it HST} WFPC2, and at F110W, F160W, and F205W of {\it HST} NICMOS 
are presented. 
The solid error bars indicate that magnitude of image B minus  
that of image A, the brightest image at F160W.}
\label{fig:twopo}
\end{figure}

A part of objects whose observed color differences are difficult 
to reproduce has been observed with more wavebands than others. 
For instance, B0218+357 and APM08279+5255 have been 
observed with 6 and 5 different wavebands, respectively. 
By using those data, we can examine the color differences
between images more intensively. 
In Figure~\ref{fig:twopo}, 
we present the magnitude differences between images.   
In the case of APM08279+5255, the trend is rather simple.  
The magnitude difference monotonically decreases with increasing wavelength, 
and bluer photons are absorbed more than redder photons at image B  
and/or are magnified more than redder photons at image A. 
As we can also see from Figure~\ref{fig:discuss}, 
the color difference can be reproduced by a  
flatter extinction law and/or a larger scale length of caustics 
than that we used in this paper. The extinction law becomes
flatter if the grain size distribution is more biased to a larger
size. Furthermore, a larger scale length of caustics is also realized 
by taking into account large variety of caustics (e.g., Appendix C), 
and thus, it is possible to reproduce observed color differences 
of APM08279+5255 either by dust extinction or by quasar microlensing.  
However, the trend is complicated in the case of B0218+357, 
because the observed color difference seems to have a peak 
around $7000 \AA$. 
If we can treat two data sets presented in Figure~\ref{fig:twopo} 
equivalently, the magnitude difference is 0.5--1.0~mag  
at $\sim 3000 \AA$ and at $\sim 16000 \AA$, but 
is $\sim 2.5 {\rm mag}$ at $\sim 7000 \AA$. 
It is clear that the magnitude difference rapidly increases 
up to a wavelength of $\sim 5000 \AA$ and becomes almost constant at
wavelengths of $\sim 5000--8000 \AA$. 
Such a wavelength dependence cannot be reproduced by dust grains. 
In contrast, quasar microlensing can magnify 
a part of quasar accretion disk selectively, 
and an emitting region of photons with a wavelength
of $\sim 7000 \AA$ can be selectively magnified with 
a certain configuration of the accretion disk and caustics.  
Further, if photons with wavelength above $\sim 7000 \AA$ 
mainly originate from more extended region than the accretion disk, 
such as a lobe of jets, a torus or a starburst region, 
these photons cannot strongly be magnified by microlensing. 
Considering this possibility of microlensing,
B0218+357 is more likely to be magnified by 
quasar microlensing than dust extinction.

As for the observational properties of 
objects whose observed color differences are not reproduced 
by either differential dust extinction or quasar microlensing, 
we cannot find any clear difference between these objects and 
the objects whose observed color differences are well reproduced 
by either differential dust extinction or quasar microlensing. 
Therefore, the peculiarity of lens galaxies or lensed quasars
is unlikely to be a probable reason for the extreme color
differences. It may be more natural to consider that 
all the observed color differences can be reproduced 
by a combination of all three possibilities presented 
in this paper without requiring any special properties of dust and/or quasar.

\subsection{Concluding remarks}

It is worth mentioning that there are some limitations 
in our current treatments of differential dust extinction 
and quasar microlensing. 

For differential dust extinction, 
we have applied empirical extinction laws 
derived for the local universe such as the MW and the SMC.  
Although the observational color differences are roughly
consistent with the local extinction curves, 
the extinction laws in distant galaxies 
such as the lens galaxies may be significantly different from the
local ones, especially in lens galaxies whose data points 
deviate significantly from our calculations in Figure~\ref{fig:difdust}.
In contrast, we have used gas distribution obtained by 
a numerical simulation for a disk galaxy \citep[see][]{hirashita}
to investigate the expected differential dust extinction.
The current sample of lens galaxies involves both early- and late-type
galaxies  (see Table~\ref{tab:gal}), which 
may have very different gas and dust properties. 
To take into account such variety of lens galaxies, 
gas and dust distribution for various types of galaxies, especially for 
early-type galaxies, and further exploration 
by using the distribution is required. 

For quasar microlensing, 
we applied a straight line caustic approximation 
and did not take into account complex caustic networks 
\citep[e.g.,][]{wam91,goicoe} that would be expected in most of systems. 
On the one hand, multiple caustics will produce more dramatic 
and/or complex structures in the expected light curves of quasar microlensing, 
and even larger or more strange features will be expected 
in the observed color difference.  
On the other hand, caustics with short scale length, i.e., small $x_s$, 
or small caustics can just slightly change the observed color 
during microlensing events. 
As already pointed out by \citet{yonehara},   
the expected color difference also 
depends on accretion disk models for quasars,  
because the spatial emissivity distribution and the wavelength dependence 
is diverse in different accretion disk models 
such as a radiatively inefficient accretion flow. 
For instance, as presented in \citet{yonehara}, 
the expected light curves of microlensing of 
an advection-dominated accretion flow (ADAF) 
is almost achromatic, at least in optical wavebands, 
since the spatial emissivity  distribution of ADAF 
is almost independent of the wavelength.
Thus, the ADAF model has difficulty in explaining 
the achromaticity of lensed quasars. 
Though the dependence on the source emissivity profile 
is less prominent in statistical properties of the light curves
\citep[][]{morton}, there exists clear dependence not only on 
the source size but also on the source emissivity profile 
in individual light curves. 
Quantitative estimates of various complexity caused by 
a variety of extinction curve, of caustic networks, 
and of accretion disks are addressed in future works.

\clearpage
\onecolumn

\begin{appendix}

\section{Extinction Law in the SMC} 

\citet{gordon} have investigated extinction curves 
of the Small Magellanic Cloud (SMC), 
the Large Magellanic Cloud, and the Milky Way 
by using photometric data in the wavelength range 
from ultraviolet to near-infrared. 
Since the coefficient $C_1$ in equation (5) of \citet{gordon}  
is negative 
\citetext{Table 3 in \citealp{gordon}} for the SMC, 
the extinction at some long wavelength becomes negative value, 
i.e., $A(x \rightarrow 0) < 0$ for positive $A_V$.  
This may not be ideal in some situation, 
and we made a fitting formulae for it by ourselves 
by using the same data for the SMC  
\citetext{denoted ``SMC Bar'' in Table 4 of \citealp{gordon}}. 

We have adopted the same functional form of the extinction curve 
as the one provided by equation 1 in \citet{cardelli}
in our fitting procedures. 
We apply the same function form as $a(x)$ in \citet{cardelli}, 
and the fit is satisfactory as shown later. 
The number of parameters to be determined is 20 in total; 
1 coefficient and 1 power index at 
$\lambda^{-1} \le 1.1~{\rm \mu m^{-1}}$, 
7 coefficients at 
$1.1~{\rm \mu m^{-1}} \le \lambda^{-1} \le 3.3~{\rm \mu m^{-1}}$,  
7 coefficients at 
$3.3~{\rm \mu m^{-1}} \le \lambda^{-1} \le 8.0~{\rm \mu m^{-1}}$, 
and 4 coefficients at 
$8.0~{\rm \mu m^{-1}} \le \lambda^{-1}$. 
We have determined those coefficients by performing  
$\chi^2$ minimization with a downhill simplex method \citep[][]{press}.

\begin{figure}
\centering
\includegraphics[width=\hsize]{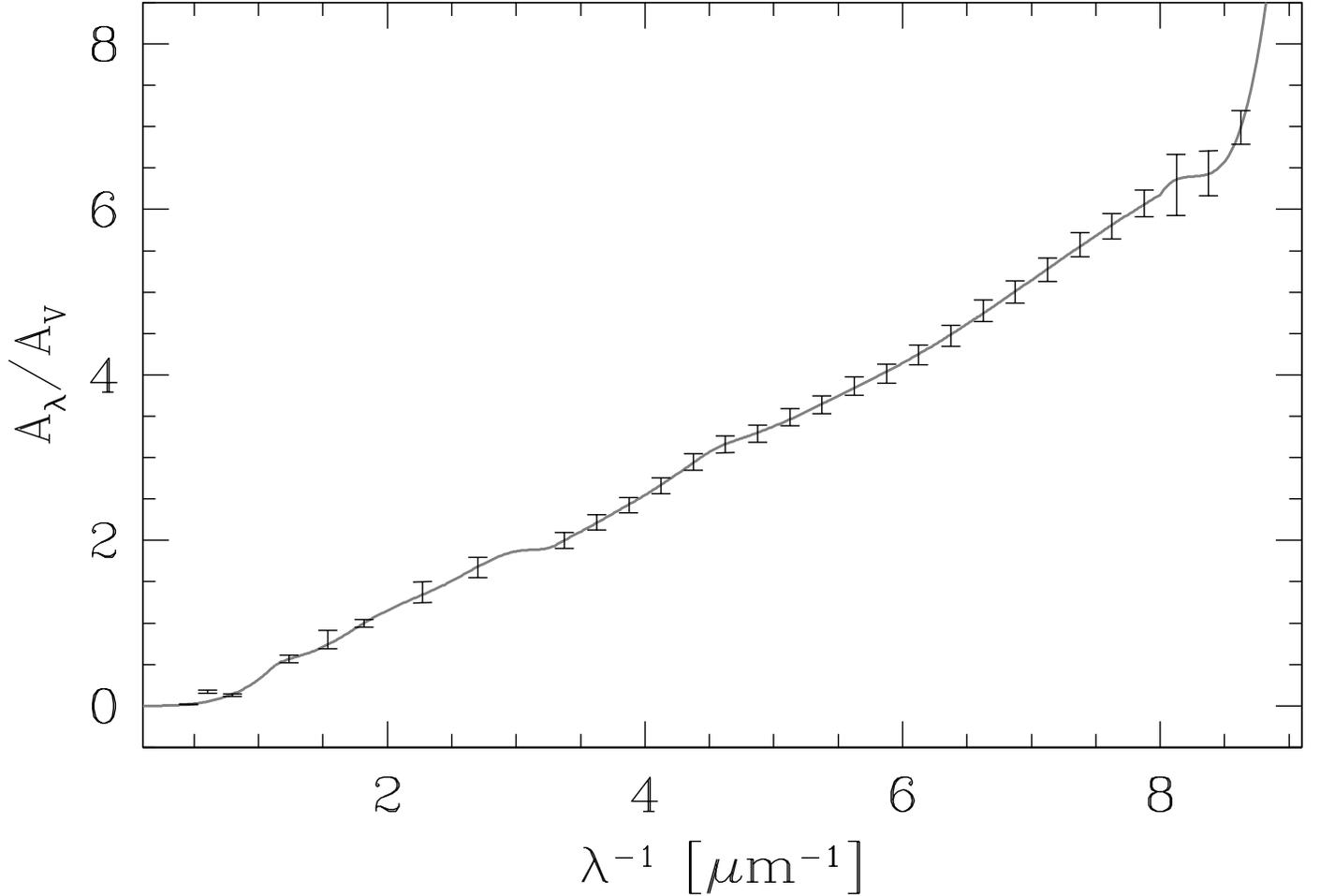}
\caption{The SMC extinction curve obtained by our fitting procedure 
is shown by the gray solid line. 
Abscissa is inverse of wavelength ($\lambda^{-1}$), 
and ordinate is extinction value divided by extinction at $V$ band 
($A_{\lambda}/A_V$).
Data points with the error used in our fitting procedure 
\citetext{originally presented by \citealp{gordon}} 
are overlaid as the black bars on the extinction curve. }
\label{fig:appendix}
\end{figure}

The resultant formulae for the extinction as a function of 
the inverse of wavelength ($x = \lambda^{-1}$ in $\mu$m$^{-1}$)
is as follows; 
\begin{eqnarray}
\frac{A(x)}{A_V} 
 &=& 0.31884x^{3.57967} ~ ~ ~ (x \le 1.1) \\
 &=& 1 + 0.92733p - 0.31777p^2 - 0.83093p^3 + 1.58876p^4 \nonumber \\ 
 &~& + 0.60622p^5 - 1.91599p^6 + 0.71769p^7~ ~ ~ (1.1 \le x \le 3.3) \\
 &=& -0.77359 + 0.81739x + 
  \frac{0.02248}{\left( x - 4.54152 \right)^2 + 0.13635} \nonumber \\ 
 &~& + 0.23260q^2 - 0.06653q^3~ ~ ~ (3.3 \le x \le 8.0) \\
 &=& 6.17700 + 2.39744r - 8.90103r^2 + 11.44564r^3~ ~ ~ (8.0 \le x),  
\end{eqnarray}
where $p = x - 1.82$, $q = x - 5.9$, and $r = x - 8.0$.  The 
$\chi^2$ value for the best fit parameters is $36.28$. 
Since the number of data points that we have used in our fitting 
is 30 and the degree of freedom becomes $30-20=10$, 
the reduced $\chi^2$ value clearly exceeds $1.0$ 
and thus is not particularly good. 
This mainly originates in a few data points with small error bars 
at longer wavelength, and we can see in Figure~\ref{fig:appendix} 
that our fitting result nicely traces most of the data points, 
especially in the wavelength range relevant in this paper. 

\section{Supplementary remarks for individual objects}

\subsection{Possible evidence for quasar microlensing}

From previous measurements, the existence of a quasar microlensing signal 
is quite obvious from their observed light curves. 
For instance, Q2237+0305 has frequently shown clear microlensing signals 
\citep[e.g.,][]{woz} in the photometric data  
since the first detection of microlensing by \citet{irw89}.
Additionally, due to the large leverage arm of the nearby lens galaxy, 
the transverse velocity of the caustics on the source plane 
is enhanced by a factor of $10$ from that of other lensed quasars, 
and the typical timescale of the microlensing events is reduced by 
the same factor. 

Possible microlensing signals in optical wavebands  
are suggested in Q0957+561 by \citet{colley,goicoe}, 
and ones in radio wavebands are suggested in B1600+434  
by \citet{koop01}.   
\citet{burud00} have also shown a possibility of quasar 
microlensing signal in the $I$-band monitoring data of B1600+434. 

For the measurement of time delay between images, 
contributions from quasar microlensing 
are included in the analysis, 
and the goodness of fit is improved substantially  
by including effects of quasar microlensing.  
From this point of view, 
the existence of quasar microlensing is indicated indirectly 
in a subset of the current sample, 
PG1115+080 \citep[][]{barka}, 
RXJ0911+0551 \citep[][]{hjorth},  
HE1104-1805 \citep[][]{ofek}, 
SBS1520+530 \citep[][]{gaynul}, 
and FBQ0951+2635 \citep[][]{jakob}. 
Observations of all these objects are performed at optical wavebands. 

For LBQS1009-0252, sparse multi-color monitoring observations 
have been done and some spectra have been obtained by \citep[][]{claesk}.
\citet[][]{claesk} have focused on the changes of colors and spectra  
during the observations and have mentioned a possible existence of 
dust extinction and/or microlensing in this system.  
However, it may be possible to explain the changes by
intrinsic variabilities. 

In the case of Q0142-100, evidence for quasar microlensing is 
considered to be obtained from the observed color changes 
in optical monitoring data \citep[][]{nakos}. 

Also in the case of HE2149-2745,  
quasar microlensing is suggested   
to explain unexpected spectral differences 
between images \citep[][]{burud02a}. 


\subsection{Other remarks}

One of the most famous lensed quasars, Q0957+561, 
is in a somewhat special environment compared to other objects, 
because the lens galaxy is obviously a member of a galaxy cluster 
\citep[e.g.,][]{ango}. 

MG2016+112 is supposed to have two lens galaxies. 
Although the lens galaxies show the brightness profile 
of early type galaxies, 
one or both of the lens galaxies show an [O\,{\sc ii}] emission line 
\citep[][]{koop02}. 

Radio monitoring observations of MG0414+0534 
have been performed by \citet{moore}. 
Unfortunately, the intrinsic flux variation of this quasar 
seems to be small, and a time delay measurement has not been successful. 

B1030+074 has been detected by \citet{xan}, and  
at the same time, they claimed that 
an image of the lens galaxy which is taken by {\it HST} 
shows substructure which could be a spiral arm 
or an interacting galaxy. 

For B2045+265, all the three images show 
almost the same color difference from, or redder than, 
the brightest image in F160W filter (see Figure~\ref{fig:discuss}). 
However, it is not probable that three of four images are
occasionally affected 
by dust extinction and/or quasar microlensing in the same way. 
Rather, it would be reasonable to consider
that only the brightest image in F160W filter 
is predominantly affected by dust extinction and/or quasar
microlensing, which cause a common color difference for
the remaining three images.

\section{Scale Length of Caustics}

To estimate color changes produced by quasar microlensing, 
we have applied a simple approximation  
in the vicinity of fold caustics shown in equation~\ref{eq:qmlappmag}. 
This approximation formula includes a scale length which is denoted 
by $x_s$, and this scale length is not arbitral but determined 
by a property of the caustics \citep[e.g.,][]{sef}, 
i.e., external convergence and shear, and distribution of lens objects. 
Here, we briefly investigate how to estimate the length, 
and we present the expected value. 

\subsection{Definition of the scale length}

By using a lens potential, $\psi$, 
the lens equation for quasar microlensing with $N$ point mass lenses 
is expressed as 
\begin{eqnarray}
0 &=& \frac{\partial \psi}{\partial \theta_x} 
 \equiv -\beta_x + (1 - \kappa_c) \theta_x  
  + \gamma (\cos 2\phi \theta_x + \sin 2\phi \theta_y) - \sum_{i=1}^{N} 
   \left( \frac{\alpha^{[i]}}{\left|{\bf \theta} - {\bf \theta^{[i]}}\right|} 
    \right)^2 (\theta_x - \theta_x^{[i]})  
\label{eq:lenseq_x} \\
0 &=& \frac{\partial \psi}{\partial \theta_y} 
 \equiv -\beta_y + (1 - \kappa_c) \theta_y
  + \gamma (\sin 2\phi \theta_x - \cos 2\phi \theta_y) - \sum_{i=1}^{N} 
   \left( \frac{\alpha^{[i]}}{\left|{\bf \theta} - {\bf \theta^{[i]}}\right|} 
    \right)^2 (\theta_y - \theta_y^{[i]}),  
\label{eq:lenseq_y}  
\end{eqnarray}
where $\beta_x$, $\beta_y$, $\theta_x$, $\theta_y$, 
$\theta_x^{[i]}$, $\theta_y^{[i]}$, $\alpha^{[i]}$,
$\kappa_c$, $\gamma$, and $\phi$ represent 
$x-$ and $y-$ coordinates of source, 
$x-$ and $y-$ coordinates of image (${\bf \theta}$), 
$x-$ and $y-$ coordinates of $i-$th lens (${\bf \theta^{[i]}}$), 
Einstein ring radius of $i-$th lens, 
convergence due to smooth matter, 
external shear and direction of the shear, respectively. 
For a practical convenience in deriving the approximation formula,  
the coordinate systems are chosen such that 
the origin of the lens plane and the source plane 
is located on critical curves and caustics, respectively, 
and $y-$axis of the source plane is orthogonal to the fold caustics. 
Performing the Taylor expansion and 
applying conditions in the vicinity of the fold caustics, 
we can obtain the following expression for 
magnification ($\mu$) after some algebra,  
\begin{equation}
\mu(\beta_x, \beta_y) \simeq \left[ \frac{1}{2}  
 \left( \left. \frac{\partial^2 \psi}{\partial \theta_x^2} 
  \right|_{(0,0)} \right)^{2}  
   \left. \frac{\partial^3 \psi}{\partial \theta_y^3} 
    \right|_{(0,0)} \beta_y \right] ^{-1/2}.  
\end{equation} 
Here, the caustics are assumed to be straight lines, 
and the dependence on $\beta_x$ disappears from the magnification. 
This is the same form as the first term in equation~\ref{eq:qmlappmag} 
for inside the caustics, and now it is clear that 
the scale length in equation~\ref{eq:qmlappmag} should be 
\begin{equation}
x_s = 2 \left[   
 \left( \left. \frac{\partial^2 \psi}{\partial \theta_x^2} 
  \right|_{(0,0)} \right)^{2}  
   \left. \frac{\partial^3 \psi}{\partial \theta_y^3} 
    \right|_{(0,0)} \right] ^{-1}.  
\end{equation} 
See \citet{sef} for more details.

\subsection{Efficient method to estimate the scale length} 

Although the magnification probability of quasar microlensing 
depends on the mass spectrum of the lens objects \citep[][]{schec04}, 
we assumed that all the lenses have the same mass.  
Further, all the length scales ($x_s$) are normalized 
to the Einstein ring radius for an assigned lens mass, 
and $\alpha^{[i]}$ is set to be unity here. 
The following estimation method can be applicable  
also if the mass spectrum is taken into account. 

In addition to equations~\ref{eq:lenseq_x} and~\ref{eq:lenseq_y}, 
the following conditions should also be satisfied on fold caustics,  
or corresponding critical curves, i.e., at $(\theta_x, \theta_y)=(0, 0)$; 
\begin{eqnarray}
0 &=& \frac{\partial^2 \psi}{\partial x \partial y} 
 \equiv \gamma \sin 2\phi + \sum_{i=1}^{N} 
  \frac{2 (\theta_x - \theta_x^{[i]}) (\theta_y - \theta_y^{[i]})}
   {\left|{\bf \theta} - {\bf \theta}^{[i]}\right|^4} 
\label{eq:fold_xy} \\
0 &=& \frac{\partial^2 \psi}{\partial y^2} 
 \equiv (1 - \kappa_c) - \gamma \cos 2\phi + \sum_{i=1}^{N} 
  \frac{(\theta_y - \theta_y^{[i]})^2 - (\theta_x - \theta_x^{[i]})^2}
   {\left|{\bf \theta} - {\bf \theta}^{[i]}\right|^4}
\label{eq:fold_yy} \\
0 &\neq& \frac{\partial^2 \psi}{\partial x^2} 
 \equiv (1 - \kappa_c) + \gamma \cos 2\phi + \sum_{i=1}^{N} 
  \frac{(\theta_x - \theta_x^{[i]})^2 - (\theta_y - \theta_y^{[i]})^2}
   {\left|{\bf \theta} - {\bf \theta}^{[i]}\right|^4}
\label{eq:fold_xx} \\
0 &\neq& \frac{\partial^3 \psi}{\partial y^3} 
 \equiv \sum_{i=1}^{N} 2(\theta_y - \theta_y^{[i]}) 
  \frac{3(\theta_x - \theta_x^{[i]})^2 - (\theta_y - \theta_y^{[i]})^2}
   {\left|{\bf \theta} - {\bf \theta}^{[i]}\right|^6}.
\label{eq:fold_yyy}
\end{eqnarray}
As we can recognize in the above equations, 
equation~\ref{eq:fold_xx} is simply reduced to 
$2(1-\kappa_c)$ by using equation~\ref{eq:fold_yy}. 
Therefore, the only term we should evaluate to estimate $x_s$ 
is the third derivative expressed in equation~\ref{eq:fold_yyy}. 
A condition for equation~\ref{eq:fold_yyy} corresponds to  
a condition to avoid cusp caustics.  

Since the above equations should be satisfied on fold caustics or 
corresponding critical curves, i.e., $(\theta_x, \theta_y)=(0, 0)$,  
equations~\ref{eq:lenseq_x},~\ref{eq:lenseq_y},~\ref{eq:fold_xy}, 
and~\ref{eq:fold_yy} can be rewritten as the following forms,  
\begin{eqnarray}
0 &=& \sum_{i=1}^{N-2} \frac{\theta_x^{[i]}}{\left|{\bf \theta^{[i]}}\right|^2} 
 + \frac{\theta_x^{[N-1]}}{\left|{\bf \theta^{[N-1]}}\right|^2} 
  + \frac{\theta_x^{[N]}}{\left|{\bf \theta^{[N]}}\right|^2} 
\label{eq:apply_x} \\
0 &=& \sum_{i=1}^{N-2} \frac{\theta_y^{[i]}}{\left|{\bf \theta}^{[i]}\right|^2} 
 + \frac{\theta_y^{[N-1]}}{\left|{\bf \theta}^{[N-1]}\right|^2}
  + \frac{\theta_y^{[N]}}{\left|{\bf \theta}^{[N]}\right|^2}
\label{eq:apply_y}  \\ 
0 &=& \gamma \sin 2\phi + \sum_{i=1}^{N-2} 
  \frac{2\theta_x^{[i]}\theta_y^{[i]}}{\left|{\bf \theta}^{[i]}\right|^4} 
   + \frac{2\theta_x^{[N-1]}\theta_y^{[N-1]}}{\left|{\bf \theta}^{[N-1]}\right|^4} 
    + \frac{2\theta_x^{[N]}\theta_y^{[N]}}{\left|{\bf \theta}^{[N]}\right|^4} 
\label{eq:apply_xy} \\
0 &=& (1 - \kappa_c) - \gamma \cos 2\phi + \sum_{i=1}^{N-2} 
  \frac{(\theta_y^{[i]})^2-(\theta_x^{[i]})^2}
   {\left|{\bf \theta}^{[i]}\right|^4}
    + \frac{(\theta_y^{[N-1]})^2-(\theta_x^{[N-1]})^2}
     {\left|{\bf \theta}^{[N-1]}\right|^4}
      + \frac{(\theta_y^{[N]})^2-(\theta_x^{[N]})^2}
       {\left|{\bf \theta}^{[N]}\right|^4}. 
\label{eq:apply_yy} 
\end{eqnarray}
Distributing the lens objects with the index from $i=1$ to $i=N-2$ randomly,  
equations~\ref{eq:apply_x},~\ref{eq:apply_y},~\ref{eq:apply_xy}, 
and~\ref{eq:apply_yy} are equivalent with 4 equations for 
4 unknown quantities ($\theta_x^{[N-1]}$, $\theta_y^{[N-1]}$, 
$\theta_x^{[N]}$, and $\theta_y^{[N]}$)  
for given $\gamma$ and $\phi$.  
By solving those 4 equations, 
we can obtain locations of two lens objects 
to create fold caustics on the origin of the applied coordinate. 
Substituting locations of all the lens objects 
including ${\bf \theta}^{[N-1]}$ and ${\bf \theta}^{[N]}$ 
into equation~\ref{eq:fold_yyy}, 
we can obtain $x_s$ for the fold caustics 
in this realization of distribution of the lens objects. 
If the number of the lens objects which are randomly distributed 
on the lens plane ($N-2$ in this case) is 
large enough compared to the number of the lens objects 
which are added to create the fold caustics at a certain place 
on the source plane ($2$ in this case),  
the resultant distribution of the lens objects 
can be also treated as a random distribution. 

An advantage of this procedure is 
that calculations of caustics networks are not required,  
and that computations for one realization are very fast. 
Although the procedure can only provide $x_s$ of caustics  
rather than the whole magnification pattern, 
it is efficient to investigate statistical properties of $x_s$. 
The procedure must be useful to evaluate a proper value of $x_s$ 
for a given environment, 
i.e., total convergence ($\kappa$), $\kappa_c$, and $\gamma$. 

\begin{figure}
\centering
\includegraphics[width=\hsize]{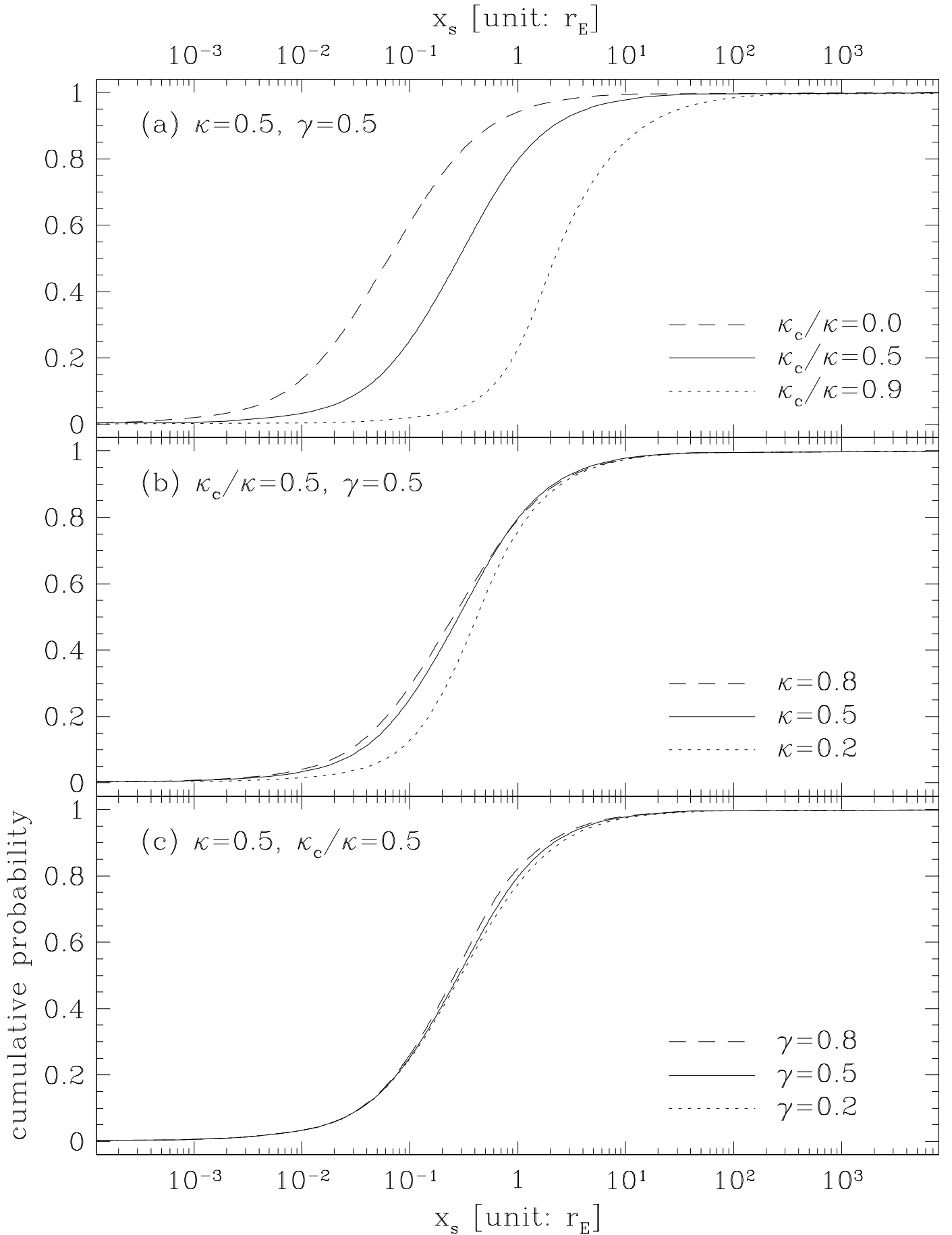}
\caption{Cumulative probability distributions of 
the scale length of fold caustics, $x_s$.  
Abscissa is $x_s$ in units of the Einstein ring radius ($r_E$).
In the top panel (panel a), total convergence ($\kappa$) and 
external shear ($\gamma$) are fixed to $0.5$. 
Distributions for a fraction of smooth matter ($\kappa_c/\kappa$) 
equals $0.0$, $0.5$, and $0.9$ are presented by 
the dashed, solid, and dotted lines, respectively.
In the middle panel (panel b), $\kappa_c/\kappa$ and $\gamma$ 
are fixed as $0.5$. 
Distributions for $\kappa=0.8$, $0.5$, and $0.2$ are presented by 
the dashed, solid, and dotted line, respectively.
In the bottom panel (panel c), $\kappa$ and $\kappa_c/\kappa$ 
are fixed as $0.5$. 
Distributions for $\gamma=0.8$, $0.5$, and $0.2$ are presented by 
the dashed, solid, and dotted line, respectively.}
\label{fig:xscale}
\end{figure}

We have estimated $x_s$ with $10^4$ lens objects (i.e., $N=10^4$) 
for several values of $\kappa$, $\kappa_c$, and $\gamma$. 
All the lens objects except the last two, $(N-1)-$th and $N-$th lens objects, 
are randomly distributed within a circle.  
Radius of the circle is determined so that the surface density of 
the lens objects is equal to $\kappa - \kappa_c$, 
i.e., $\left[ N/(\kappa - \kappa_c) \right]^{1/2}$. 
Direction of external shear, $\phi$, is also chosen randomly 
at every realization. 
We have performed $10^5$ realizations to obtain 
the expected distributions of $x_s$,  
and results are presented in Figure~\ref{fig:xscale}. 

First, the distribution depends on the fraction of smooth matter 
with respect to the total convergence, $\kappa_c/\kappa$. 
The scale length $x_s$ becomes smaller as $\kappa_c/\kappa$ becomes smaller. 
Since $\kappa_c/\kappa$ cannot be smaller than $0$, 
the dashed line in Figure~\ref{fig:xscale}(a) 
can be treated as a minimum of $x_s$ distribution 
for given $\kappa$ and $\gamma$. 
Although it is practically difficult to investigate $\kappa_c/\kappa$, 
we obtain the minimum of $x_s$ for $\kappa=0.5$ and $\gamma=0.5$ 
as $\sim 10^{-3}$ from Figure~\ref{fig:xscale}(a)  
\footnote{The results may show a different aspect of 
what \citet{schec04} found.  
However, the distribution obtained here cannot be simply transformed 
into magnification probabilities, because we did not take into account 
neither spatial distribution of caustics nor size of caustics.}.   
Second, as is shown in Figure~\ref{fig:xscale}(b) and (c), 
dependence on $\kappa$ and $\gamma$ is weak.   
The minimum value of $x_s$ increases slightly with decreasing $\kappa$, 
and the maximum value of $x_s$ increases slightly with decreasing $\gamma$. 
Especially, the dependence of the $x_s$ distribution on $\gamma$ 
is almost negligible. 
Although the distribution of $x_s$ shows a slight dependence on 
$\kappa_c/\kappa$, $\kappa$, and $\gamma$, 
we can conclude that typical value of $x_s$ is within 
a range of 0.1 -- 1 as shown in Figure~\ref{fig:xscale}. 
Several scale lengths at certain cumulative probabilities 
are also summarized in Table~\ref{tab:xs} for convenience.

\begin{table*}
\caption{The scale lengths of caustics, $x_s$, 
at certain cumulative probabilities for distributions 
presented in Figure~\ref{fig:xscale} are listed. 
The column for $50\%$ indicates the median value of the distributions. 
$15.85\%$ and $84.15\%$ indicate the $\pm 1\sigma$ confidence levels  
from the median value, and $0.135\%$ and $99.865\%$ indicate 
the $\pm 3\sigma$ confidence levels from the median value.} 
\label{tab:xs}
\centering
\begin{tabular}{c c c | c c c c c} 
\hline \hline
 $\kappa$ & $\kappa_c/\kappa$ & $\gamma$ & $0.135\%$ & $15.85\%$ & $50\%$ & $84.15\%$ & $99.865\%$ \\
\hline
 0.5 & 0.5 & 0.5 & $9.86 \times 10^{-5}$ & $5.79 \times 10^{-2}$ & $2.83 \times 10^{-1}$ & $1.33$ & $1.61 \times 10^2$  \\
 0.5 & 0.9 & 0.5 & $2.04 \times 10^{-3}$ & $7.51 \times 10^{-1}$ & $2.20$ & $9.23$ & $1.29 \times 10^3$  \\
 0.5 & 0.0 & 0.5 & $1.96 \times 10^{-5}$ & $1.18 \times 10^{-2}$ & $6.37 \times 10^{-2}$ & $3.42 \times 10^{-1}$ & $4.80 \times 10^1$  \\
\hline
 0.2 & 0.5 & 0.5 & $2.70 \times 10^{-4}$ & $1.21 \times 10^{-1}$ & $4.18 \times 10^{-1}$ & $1.57$ & $1.65 \times 10^2$ \\
 0.8 & 0.5 & 0.5 & $7.62 \times 10^{-5}$ & $4.63 \times 10^{-2}$ & $2.53 \times 10^{-1}$ & $1.38$ & $1.82 \times 10^2$ \\
\hline
 0.5 & 0.5 & 0.2 & $9.86 \times 10^{-5}$ & $5.84 \times 10^{-2}$ & $2.98 \times 10^{-1}$ & $1.52$ & $2.06 \times 10^2$ \\
 0.5 & 0.5 & 0.8 & $9.85 \times 10^{-5}$ & $5.66 \times 10^{-2}$ & $2.59 \times 10^{-1}$ & $1.15$ & $1.44 \times 10^2$ \\
\hline
\end{tabular}
\end{table*}

\end{appendix}

\clearpage
\twocolumn

\begin{acknowledgements}
We acknowledge J. Wambsgan{\ss}, E. Mediavilla, R. Schmidt, A. Cassan, 
E. Koptelova, C. Faure, T. Anguita, 
J. Fohlmeister, and M. Zub for their valuable comments 
and encouragements, 
and an anonymous referee for her/his nice suggestions.
We thank the Japan-Italy seminar which is supported jointly 
by the Japan Society for the Promotion Science and 
by the National Research Council of Italy. 
A.Y. also acknowledges the Japan Society for the Promotion 
of Science (09514) and Inoue Foundation for Science.
H.H. has been supported by Grants-in-Aid for Scientific
Research of the Ministry of Education, Culture, Sports,
Science and Technology (Nos. 18026002 and 18740097). 
P.R. acknowledges financial support by the German
\emph{Deut\-sche For\-schungs\-ge\-mein\-schaft}, DFG,
through Emmy-Noether grant Ri 1124/3-1. 

\end{acknowledgements}

\clearpage
\onecolumn

\tablecaption{Current sample of lensed quasars. 
The redshifts of the lens galaxy ($z_l$) and of the source quasar
($z_s$), the colors of the lensed quasar images and the errors 
for F555W - F160W ($m_{F555W} - m_{F160W}$), 
and for F814W - F160W ($m_{F814W} - m_{F160W}$) are presented. 
These colors should be identical for each image of 
the same lensed quasar unless intrinsic variability, dust extinction, 
or microlensing play a significant role.  
The magnitudes of the images at F160W ($m_{F160W}$) are also presented 
in the last column.  
The redshifts of the lens galaxies are not known for 
the 6 objects listed at the bottom. 
All the data are taken from the CASTLES web page, 
and the colors are calculated from these data. } 
\label{tab:obj} 
\begin{center}
\setcounter{table}{1}
\begin{supertabular}{r c c c c c}
\hline \hline
 Object Name & $z_l$ & $z_s$ & $m_{F555W} - m_{F160W}$ & $m_{F814W} - m_{F160W}$ & $m_{F160W}$ \\
\hline
Q0142-104~A  & 0.49 & 2.72 &  $1.61 \pm 0.092$ & $1.34 \pm 0.181$ & $15.28 \pm 0.02$ \\
         ~B  & ~    & ~    &  $1.56 \pm 0.050$ & $1.14 \pm 0.153$ & $17.57 \pm 0.03$ \\
B0218+357~A  & 0.68 & 0.96 &  $5.76 \pm 0.036$ & $4.31 \pm 0.073$ & $17.52 \pm 0.02$ \\
         ~B  & ~    & ~    &  $4.17 \pm 0.022$ & $2.45 \pm 0.028$ & $16.94 \pm 0.02$ \\
MG0414+0534~A1 & 0.96 & 2.64 &  $9.99 \pm 1.270$ & $4.89 \pm 0.063$ & $15.54 \pm 0.02$ \\
           ~A2 & ~    & ~    & $10.82 \pm 0.272$ & $5.49 \pm 0.067$ & $15.87 \pm 0.03$ \\
           ~B  & ~    & ~    &  $9.52 \pm 2.080$ & $4.68 \pm 0.076$ & $16.56 \pm 0.03$ \\
           ~C  & ~    & ~    &  $9.27 \pm 0.141$ & $4.69 \pm 0.073$ & $17.41 \pm 0.02$ \\
B0712+472~A  & 0.41 & 1.34 &  $3.78 \pm 0.277$ & $2.64 \pm 0.434$ & $20.46 \pm 0.21$ \\
         ~B  & ~    & ~    &  $4.58 \pm 0.231$ & $3.14 \pm 0.422$ & $21.08 \pm 0.22$ \\
         ~C  & ~    & ~    &  $5.25 \pm 0.272$ & $3.32 \pm 0.467$ & $21.05 \pm 0.08$ \\
         ~D  & ~    & ~    &  $4.77 \pm 0.940$ & $2.97 \pm 0.433$ & $21.90 \pm 0.14$ \\
RXJ0911+0551~A  & 0.77 & 2.80 &  $1.24 \pm 0.045$ & $0.79 \pm 0.045$ & $17.59 \pm 0.02$ \\
            ~B  & ~    & ~    &  $1.53 \pm 0.054$ & $0.99 \pm 0.036$ & $17.65 \pm 0.02$ \\
            ~C  & ~    & ~    &  $1.55 \pm 0.067$ & $1.02 \pm 0.050$ & $18.34 \pm 0.03$ \\
            ~D  & ~    & ~    &  $1.62 \pm 0.102$ & $1.01 \pm 0.036$ & $18.65 \pm 0.02$ \\
SBS0909+523~A  & 0.83 & 1.38 &  $2.53 \pm 0.082$ & $1.95 \pm 0.028$ & $14.60 \pm 0.02$ \\
           ~B  & ~    & ~    &  $3.46 \pm 0.173$ & $2.20 \pm 0.222$ & $14.73 \pm 0.03$ \\
BRI0952-0115~A & 0.63 & 4.50 & $3.59 \pm 0.036$ & $1.46 \pm 0.045$ & $17.07 \pm 0.02$ \\
            ~B & ~   & ~    & $3.55 \pm 0.036$ & $1.49 \pm 0.045$ & $18.44 \pm 0.02$ \\ 
Q0957+561~A  & 0.36 & 1.41 &  $1.49 \pm 0.085$ & $1.11 \pm 0.104$ & $15.60 \pm 0.03$ \\
         ~B  & ~    & ~    &  $1.43 \pm 0.067$ & $1.10 \pm 0.095$ & $15.68 \pm 0.03$ \\
LBQS1009-0252~A  & 0.87 & 2.74 &  $1.75 \pm 0.082$ & $1.32 \pm 0.036$ & $16.63 \pm 0.02$ \\
             ~B  & ~    & ~    &  $2.97 \pm 0.089$ & $1.79 \pm 0.057$ & $18.20 \pm 0.04$ \\
B1030+071~A  & 0.60 & 1.54 &  $4.82 \pm 0.221$ & $3.59 \pm 0.250$ & $15.88 \pm 0.02$ \\
         ~B  & ~    & ~    &  $3.83 \pm 0.212$ & $2.59 \pm 0.242$ & $19.79 \pm 0.03$ \\
HE1104-1805~A  & 0.73 & 2.32 &  $1.35 \pm 0.321$ & $0.81 \pm 0.050$ & $15.57 \pm 0.03$ \\
           ~B  & ~    & ~    &  $1.66 \pm 0.165$ & $0.97 \pm 0.064$ & $17.04 \pm 0.04$ \\
PG1115+080~A1 & 0.31 & 1.72 &  $1.19 \pm 0.112$ & $0.71 \pm 0.028$ & $15.71 \pm 0.02$ \\
          ~A2 & ~    & ~    &  $1.41 \pm 0.095$ & $0.64 \pm 0.042$ & $16.21 \pm 0.03$ \\
          ~B  & ~    & ~    &  $0.69 \pm 0.453$ & $0.21 \pm 0.393$ & $17.70 \pm 0.05$ \\
          ~C  & ~    & ~    &  $1.72 \pm 0.322$ & $1.14 \pm 0.342$ & $17.23 \pm 0.04$ \\
B1422+231~A  & 0.34 & 3.62 &  $2.02 \pm 0.112$ & $1.47 \pm 0.082$ & $14.41 \pm 0.02$ \\
         ~B  & ~    & ~    &  $2.16 \pm 0.104$ & $1.56 \pm 0.076$ & $14.29 \pm 0.03$ \\
         ~C  & ~    & ~    &  $2.11 \pm 0.076$ & $1.43 \pm 0.050$ & $14.98 \pm 0.03$ \\
         ~D  & ~    & ~    &  $2.30 \pm 0.063$ & $1.54 \pm 0.045$ & $18.14 \pm 0.02$ \\
SBS1520+530~A  & 0.72 & 1.86 &  $1.85 \pm 0.161$ & $0.77 \pm 0.036$ & $17.20 \pm 0.02$ \\
           ~B  & ~    & ~    &  $1.45 \pm 0.163$ & $0.96 \pm 0.095$ & $18.03 \pm 0.03$ \\
B1600+434~A  & 0.41 & 1.59 &  $2.95 \pm 0.614$ & $1.26 \pm 0.170$ & $20.66 \pm 0.13$ \\
         ~B  & ~    & ~    &  $1.85 \pm 0.891$ & $0.92 \pm 0.146$ & $20.47 \pm 0.14$ \\
MG2016+112~A  & 1.01 & 3.27 &  $2.44 \pm 0.289$ & $1.97 \pm 0.086$ & $20.48 \pm 0.07$ \\
          ~B  & ~    & ~    &  $3.15 \pm 0.054$ & $2.12 \pm 0.071$ & $20.50 \pm 0.05$ \\
          ~C  & ~    & ~    &  $4.23 \pm 0.711$ & $2.94 \pm 0.064$ & $20.21 \pm 0.04$ \\
B2045+265~A  & 0.87 & 1.28 &  $6.07 \pm 0.210$ & $3.14 \pm 0.110$ & $19.81 \pm 0.01$ \\
         ~B  & ~    & ~    &  $6.03 \pm 0.072$ & $3.22 \pm 0.081$ & $20.37 \pm 0.04$ \\
         ~C  & ~    & ~    &  $5.99 \pm 0.233$ & $3.13 \pm 0.126$ & $20.05 \pm 0.04$ \\
         ~S  & ~    & ~    &  $5.44 \pm 0.057$ & $2.39 \pm 0.072$ & $19.41 \pm 0.04$ \\
HE2149-2745~A  & 0.50 & 2.03 &  $1.30 \pm 0.085$ & $0.85 \pm 0.042$ & $15.67 \pm 0.03$ \\
           ~B  & ~    & ~    &  $1.44 \pm 0.076$ & $0.85 \pm 0.036$ & $17.23 \pm 0.03$ \\
Q2237+0305~A  & 0.04 & 1.69 &  $2.39 \pm 0.108$ & $0.96 \pm 0.134$ & $14.96 \pm 0.06$ \\
          ~B  & ~    & ~    &  $2.01 \pm 0.120$ & $1.75 \pm 0.110$ & $15.46 \pm 0.01$ \\
          ~C  & ~    & ~    &  $2.85 \pm 0.101$ & $1.06 \pm 0.120$ & $15.71 \pm 0.01$ \\
          ~D  & ~    & ~    &  $3.01 \pm 0.133$ & $1.39 \pm 0.050$ & $16.00 \pm 0.03$ \\
\hline
QJ0158-4325~A & --- & 1.29 & $1.63 \pm 0.133$ & $1.34 \pm 0.361$ & $16.47 \pm 0.03$ \\
           ~B & ~   & ~    & $1.64 \pm 0.173$ & $1.35 \pm 0.331$ & $17.27 \pm 0.03$ \\
APM08279+5255~A & --- & 3.87 & $2.50 \pm 0.036$ & $0.79 \pm 0.054$ & $14.46 \pm 0.02$ \\
             ~B & ~   & ~    & $4.10 \pm 0.036$ & $2.00 \pm 0.028$ & $13.64 \pm 0.02$ \\
FBQ0951+2635~A & --- & 1.24 & $1.67 \pm 0.204$ & $1.08 \pm 0.057$ & $15.62 \pm 0.04$ \\
            ~B & ~   & ~    & $1.33 \pm 0.076$ & $0.90 \pm 0.042$ & $16.99 \pm 0.03$ \\
Q1017-207~A & --- & 2.55 & $1.77 \pm 0.042$ & $1.26 \pm 0.036$ & $15.66 \pm 0.03$ \\
         ~B & ~   & ~    & $1.77 \pm 0.139$ & $1.26 \pm 0.058$ & $17.81 \pm 0.05$ \\
Q1208+101~A & --- & 3.80 & $2.51 \pm 0.124$ & $1.31 \pm 0.042$ & $15.91 \pm 0.03$ \\
         ~B & ~   & ~    & $2.11 \pm 0.045$ & $1.08 \pm 0.092$ & $17.55 \pm 0.02$ \\
FBQ1633+3134~A & --- & 1.52 & $1.77 \pm 0.163$ & $1.07 \pm 0.042$ & $15.78 \pm 0.03$ \\
            ~B & ~   & ~    & $1.72 \pm 0.151$ & $1.06 \pm 0.022$ & $17.23 \pm 0.02$ \\
\hline
\end{supertabular}
\end{center}

\end{document}